\documentclass[twocolumn,showpacs,preprintnumbers,amsmath,amssymb,nofootinbib]{revtex4-1}
\usepackage{epsfig} 
\usepackage{graphicx}% Include figure files
\usepackage{dcolumn}% Align table columns on decimal point
\usepackage{bm}% bold math
\usepackage{amsfonts}
\usepackage[usenames]{color}

\def\b{\begin{equation}}
\def\e{\end{equation}}
\def\be{\begin{eqnarray}}
\def\ee{\end{eqnarray}}

\newcommand{\insertplot}[5]{\begin{figure}
 \hfill\hbox to 0.05in{\vbox to #5in{\vfill
 \inputplot{#1}{#4}{#5}}\hfill}
 \hfill\vspace{-.1in}
 \caption{#2}\label{#3}
 \end{figure}}
 \newcommand{\inputplot}[3]{
 \special{ps: plotfile #1}
}
\newcounter{fig}

\begin{document}

\title{Acoustic clouds: \\ standing sound waves around a black hole analogue}

\author{Carolina L. Benone}
\email{lben.carol@gmail.com}
\affiliation{Faculdade de F\'{\i}sica, Universidade Federal do Par\'a, 66075-110, Bel\'em, Par\'a, Brazil}

\author{Lu\'{\i}s C. B. Crispino}
\email{crispino@ufpa.br}
\affiliation{Faculdade de F\'{\i}sica, Universidade Federal do Par\'a, 66075-110, Bel\'em, Par\'a, Brazil}

\author{Carlos Herdeiro}
\email{herdeiro@ua.pt}
\author{Eugen Radu}
\email{eugen.radu@ua.pt}
\affiliation{\vspace{2mm}Departamento de F\'\i sica da Universidade de Aveiro and CIDMA \\
Campus de Santiago, 3810-183 Aveiro, Portugal \vspace{1mm}}%

\date{January 2015}

\begin{abstract}
Under certain conditions sound waves in fluids experience an acoustic horizon with analogue properties to those of a black hole event horizon. In particular, a draining bathtub-like model can give rise to a rotating acoustic horizon and hence a rotating black hole (acoustic) analogue. We show that sound waves, when enclosed in a cylindrical cavity,  can form stationary waves around such rotating acoustic holes. These acoustic perturbations display similar properties to the scalar clouds that have been studied around Kerr and Kerr-Newman black holes; thus they are dubbed \textit{acoustic clouds}. We make the comparison between scalar clouds around Kerr black holes and acoustic clouds around the draining bathtub explicit by studying also the properties of scalar clouds around Kerr black holes enclosed in a cavity. Acoustic clouds suggest the possibility of testing, experimentally, the existence and properties of black hole clouds, using analog models.
\end{abstract}

\pacs{04.50.-h, 04.50.Kd, 04.20.Jb}
\maketitle

%%%%%%%%%%%%%%%%%%%%%%%%%%%%%%%%%%%%%%%%%%%%%%%%%%%%%%%%%%%%%%%%%%%%%%%%%%%%%%
\section{Introduction}
In a renowned work, Unruh showed that sound waves in an inviscid and irrotational fluid flow are governed by the Klein-Gordon equation on an effective geometry, called effective \textit{acoustic} spacetime~\cite{Unruh:1980cg}. He observed, moreover, that the effective geometry can be analogous to a black hole (BH) spacetime: if the fluid flow  becomes supersonic, the surface separating subsonic and supersonic flows is perceived as a one-way membrane by sound waves. Indeed, this surface is an event horizon of the effective acoustic spacetime. 

Building on Unruh's work, a variety of analogue models has been developed (see, $e.g.$, \cite{Barcelo:2005fc, CCLOV:proc}) with the promise of constructing BH analogues in the laboratory that could be used to test the physical properties of BHs. It is therefore of interest to understand if a given property observed in BH spacetimes can be mimicked by analogue models. 
Within this context, BH properties like absorption~\cite{Crispino:2007zz, Oliveira:2010zzb}, scattering~\cite{DOC2009, DO}, quasinormal modes and Regge poles~\cite{BCL, Cardoso:2004fi, DOC2010, DOC2012,Lemos:2013yoa} have been investigated in acoustic analogue spacetimes over the past few years.

Another such BH property, which has been the subject of recent studies, is the existence of stationary bound states of a massive scalar field, $\Phi$, around Kerr (or Kerr-Newman) BHs~\cite{Hod:2012px,Hod:2013zza,Herdeiro:2014goa,Hod:2014baa,Benone:2014ssa}: \textit{scalar clouds}.\footnote{See also \cite{Degollado:2013eqa,Sampaio:2014swa} for \textit{marginal} charged scalar and Proca clouds around Reissner-Nordstr\"om BHs.} These clouds exist for monochromatic modes with frequency $\omega$ and azimuthal harmonic index $m$, when the condition 
\begin{equation}
\omega=\omega_c\equiv m\Omega_H %\ ,
\label{sync}
\end{equation}
 holds, where $\Omega_H$ is the horizon angular velocity.\footnote{In the Kerr-Newman case, for a charged scalar field with charge $q$, this condition is modified to $\omega=m\Omega_H+q\Phi_H$, where $\Phi_H$ is the horizon electrostatic potential.} Consequently, these clouds are interpreted as zero modes of the superradiant instability of Kerr BHs, that amplifies modes obeying $\omega<m\Omega_H$~\cite{Press:1972zz}. Alternatively, one may regard this condition as requiring that a field mode of type $\Phi\sim e^{-i\omega t}e^{im \phi}$, using standard Boyer-Lindquist coordinates, is preserved by the horizon null generator $k=\partial_t+\Omega_H\partial_\phi$. Physically, this guarantees the absence of scalar field flux through the horizon, a necessary requirement for  the existence of stationary bound states around the BH.

Besides being of interest on their own, as equilibrium states of a matter field around a BH, scalar clouds have been related to the existence of Kerr BHs with scalar hair~\cite{Herdeiro:2014goa} (see also~\cite{Herdeiro:2014jaa,Herdeiro:2015gia}): backreacting clouds generate a new family of solutions to the Einstein-(massive)-Klein-Gordon system, continuously connected to the Kerr family. This suggests that whenever clouds of a given matter field can be found around a BH, in a linear analysis and with a stationary and axi-symmetric energy-momentum tensor, there exists a fully non-linear solution of new hairy BHs corresponding to making these clouds heavy and thus deforming the geometry~\cite{Herdeiro:2014ima}. A more fundamental mechanism for the existence of hairy BHs, however, that does not require the existence of clouds in a linear analysis but \textit{does} require the zero flux condition $\omega=m\Omega_H$ to hold, may be at place, as illustrated by the examples in~\cite{Brihaye:2014nba,Herdeiro:2014pka}.

In this paper we will show that there are cloud-like configurations around BH analogue models. Specifically, we shall be interested in acoustic holes and a [(2+1) dimensional] draining bathtub model for which the acoustic horizon is rotating~\cite{Visser:1997ux}. Acoustic perturbations of this model are described by a massless Klein-Gordon equation, unlike the aforementioned scalar clouds which occur for a massive Klein-Gordon field. The role of the mass is to provide the necessary confinement for bound states to exist. In a realistic experiment with a draining bathtub the ``spacetime'' cannot extend forever, and will be enclosed in a cavity, which we assume to be cylindrical as to respect the isometries of the background. Such cavity implements a boundary condition that replaces the effect of the mass and, when the analogue of condition (\ref{sync}) holds, produces standing wave-like bound states, which we call \textit{acoustic clouds}.

To make the comparison between these acoustic clouds and scalar clouds around a Kerr BH sharper we shall also study scalar clouds of the \textit{massless} Klein-Gordon equation for a Kerr BH enclosed in a cavity. Here the cavity is spheroidal ($r=$ constant) in Boyer-Lindquist coordinates, and two different boundary conditions for the scalar field are implemented therein. We will show that the variation of the clouds properties with the mirror location and `quantum numbers' labelling the cloud is qualitatively similar for both acoustic and scalar clouds. Let us remark that superradiance of scalar fields around BHs in a cavity has been studied in the past, for both Kerr~\cite{Press:1972zz,Cardoso:2004nk,Dolan:2012yt} and Reissner-Nordstr\"om geometries~\cite{Herdeiro:2013pia,Hod:2013fvl,Degollado:2013bha}. 

This paper is organized as follows. In Section~\ref{sec_kerr} we analyse massless scalar clouds around Kerr BHs in a cavity. We will see that their properties are reminiscent of the massive case (without mirror). This sets a reference frame for addressing the acoustic clouds in a draining bathtub model, which is done in Section~\ref{sec_acoustic}. In both Sections~\ref{sec_kerr} and~\ref{sec_acoustic}, computations are done using numerical methods. We shall briefly comment on the (un)likelihood of describing acoustic clouds analytically. We close in Section~\ref{sec_conclusions} with some final remarks.
 
We assume throughout the paper $c=c_S=\hbar =G=1$, where $c_S$ is the speed of sound.

%%%%%%%%%%%%%%%%%%%%%%%%%%%%%%%%%%%%%%%%%%%%%%%%%%%%%%%%%%%%%%%%%%%%%%%%%%%%%%
\section{Scalar clouds around Kerr BHs}
\label{sec_kerr}
%%%%%%%%%%%%%%%%%%%%%%%%

%%%%%%%%%%%%%%%%%%%%%%%%
\subsection{Background and scalar equation}
%%%%%%%%%%%%%%%%%%%%%%%%
Let us first consider the Kerr background, described by the line element
\be
{ds}^2&\ & = -\frac{\Delta}{\rho^2}(dt-a\sin^2{\theta}d\phi)^2 + \frac{\rho^2}{\Delta}dr^2\nonumber\\
 &+& \rho^2 d\theta^2 + \frac{\sin^2{\theta}}{\rho^2}[(r^2+a^2)d\phi-a dt]^2,
\ee
with
\b
\rho^2 \equiv  r^2 + a^2 \cos^2{\theta}, \hspace{0.5in} \Delta \equiv r^2 - 2Mr+a^2.
\e
$M$ is the ADM mass and $J=aM$ is the ADM angular momentum of the BH.

To solve the Klein-Gordon equation for a \textit{massless} scalar field, $\square \Phi = 0$, on this background, we make a harmonic/Fourier decomposition of the scalar field, which is therefore considered to be a sum of modes of the type $\Phi_{lm} = e^{i(m\phi - \omega t)}S_{lm}(\theta) R_{lm}(r)$. Taking $S_{lm}$ to be spheroidal harmonics, obeying

\begin{widetext}
\b
\frac{1}{\sin{\theta}}\frac{d}{d\theta}\left(\sin{\theta}\frac{d S_{lm}}{d \theta}\right)+\left(K_{lm}-a^2 \omega^2 + a^2\omega^2\cos^2{\theta}-\frac{m^2}{\sin^2{\theta}}\right)S_{lm} =0 \ , 
\label{esh}
\e
where $K_{lm}$ are separation constants, the above decomposition of the scalar field permits a full separation of variables. The radial functions $R_{lm}(r)$ obey the radial equation
\b
\Delta \frac{d}{d r}\left(\Delta \frac{d R_{lm}}{d r}\right) + \left[H^2 + (2ma\omega - K_{lm})\Delta \right]R_{lm}=0 \ ,
\label{erk}
\e
\end{widetext}
where $H\equiv (r^2+a^2)\omega-am$. We can rewrite Eq. (\ref{erk}) using the tortoise coordinates, defined by
\b
\frac{dr_*}{dr} \equiv  \frac{r^2+a^2}{\Delta} \ ,
\label{tor}
\e
and obtain a new radial equation without the first derivative term,
\begin{widetext}
\b
\frac{d^2 U_{lm}}{dr_*^2} + \left\{\frac{[H^2+ (2ma\omega - K_{lm})\Delta]}{(r^2+a^2)^2}-\frac{\Delta(\Delta+2r(r-M))}{(r^2+a^2)^3}+ \frac{3r^2\Delta^2}{(r^2+a^2)^4}\right\}U_{lm}=0 \ ,
\label{ert}
\e
\end{widetext}
with
\b
U_{lm}\equiv R_{lm}\sqrt{r^2+a^2} \ . 
\e

We are now able to analyse the behaviour of the radial solution close and far away from the BH. In the limit $r \rightarrow r_+$ we require a purely ingoing wave, while in the asymptotic case we have an outgoing wave:
\b
R_{lm}(r) \approx 
\left\{ 
\begin{array}{ll}
e^{-i(\omega-\omega_c) r_*}, \quad &\mbox{for $r\rightarrow r_+$}\ ,\\
e^{i\omega r_*}, \quad &\mbox{for $r\rightarrow \infty$}\ .
\end{array}
\right.
\label{solk}
\e
%where $\omega_c = m \Omega_c= m \frac{a}{r^2+a^2}$ is the critical frequency.
With the solutions given by (\ref{solk}) no stationary solutions exist. To have such solutions we shall impose, besides $\omega=\omega_c$, a different boundary condition, as we shall now describe.

%%%%%%%%%%%%%%%%%
\subsection{Numerical procedure and results}
\label{nprkerr}
%%%%%%%%%%%%%%%%%
To find clouds, $i.e.$ bound state like solutions, we enclose the system in a cavity whose boundary is located at 
\b
r=r_0 \ .
\e
We impose two different boundary conditions to our problem: The first is of Dirichlet type such that the scalar field vanishes at the cavity's boundary, $R_{lm}(r_0) = 0$. The second is a Neumann boundary condition such that the derivative of the field vanishes at the cavity's boundary, $d R_{lm}(r_0)/dr = 0$. These, together with condition (\ref{sync}) allow us to find massless scalar clouds around Kerr BHs. The clouds are labelled by three `quantum' numbers: $l$, the spheroidal harmonic index, $m$ the azimuthal harmonic index and $n$, the node number of the radial function.

To obtain the radial function, we shall integrate Eq.~(\ref{erk}) numerically. For that purpose, we consider the following series expansions for the separation constants $K_{lm}$
\b
K_{lm} - a^2 \omega^2 = l(l+1)+\sum_{k=1}^\infty c_k a^{2k}\omega^{2k}\ ,
\e
where the coefficients $c_k$ are given in \cite{abramowitz+stegun}, and to initiate the radial integration we take the following series expansion 
\b
\label{rah}
 R_{lm}=R_0\bigg(1+\sum_{k\geq 1}R_k(r-r_+)^k\bigg) \ ,
%&\ & \psi_{m}=\psi_0\bigg(1+\sum_{k\geq 1}\psi_k(r-r_+)^k\bigg)
\e
close to the horizon. To find the coefficients $R_k$ we substitute Eq.~(\ref{rah}) in Eq.~(\ref{erk}) and expand it in terms of $(r-r_+)$, taking $R_0=1$. We end up with a shooting problem for the radial function, using the shooting parameter $a$. Fixing the quantum numbers ($n, l$ and $m$) and using the event horizon $r_+$ as a scaling factor, we choose values for $r_0$, solve the radial equations from very close to the event horizon to the position of the mirror and find the values of $a$ that obey the appropriate boundary condition.

In Fig. \ref{fr0} we display the mirror location $r_0$ in terms of the horizon angular velocity $\Omega_H$, both scaled to the horizon radius, varying different cloud quantum numbers, for both Dirichlet boundary conditions (left panels) and Neumann boundary conditions (right panels). The first observation is that, generically, $r_0/r_+ \rightarrow \infty$, $\Omega_H \rightarrow 0$. This illustrates the fact that there are no massless clouds for Schwarzschild BHs in a cavity. Indeed this is expected from the known results of the massive case: there are no (massive) scalar clouds for the Schwarzschild BH, even though it is possible to have arbitrarily long-lived quasi-bound states \cite{Barranco:2012qs}. The second generic observation is that, as we approach the mirror to the BH horizon we need to increase $\Omega_H$ in order to have clouds. Conversely, and taking precisely the extremal case ($\Omega_H r_+ = 0.5$), we observe that increasing the value of $m=l$ the position of the mirror tends to the horizon. We shall come back to this point shortly.

The dependence on the cloud's quantum numbers observed in Fig.~\ref{fr0} can be summarized as follows.  Increasing either the spheroidal harmonic index $l$ or the node number $n$, implies that, for fixed mirror position $r_0/r_+$, clouds exist for larger horizon angular velocity $\Omega_H r_+$. The same behaviour is observed, on the other hand, if one decreases simultaneously $l=m$. All these trends are analogous to those observed for massive scalar clouds (without mirror) in~\cite{Benone:2014ssa}. They can be heuristically  interpreted in terms of a mechanical equilibrium between the BH-cloud gravitational attraction and angular momentum driven repulsion due to the BH-cloud energy currents, \textit{cf.}~\cite{Benone:2014ssa}. 

Finally concerning the differences between Dirichlet and Neumann boundary conditions, Fig.~\ref{fr0} shows that both cases present the same qualitative behaviour, but that for the same angular velocity $\Omega_H$, clouds with Neumann boundary conditions occur for a smaller $r_0$. This is quite natural, as one can always obtain a Neumann cloud from a Dirichlet cloud by decreasing the cavity where the latter stands to the point where the radial derivative of the radial function vanishes, which occurs inside the cavity.

Fig. \ref{frn} illustrates the radial profile of the massless scalar clouds for Dirichlet (left panels) and Neumann (right panels) boundary conditions for $l=m=1$ (top panels) and $l=m=2$ (bottom panels). For all cases we choose $r_0=20 r_+$ and show three distinct solutions corresponding to different numbers of nodes $n=0,1$ and $2$. The radial profile has the typical form of standing waves with fixed boundary conditions. 

Fig. \ref{fok} shows the position of the clouds $r_{MAX}$, defined as the maximum of the function $4 \pi r^2 |R_{lm}|^2$, compared with the radial value of the circular null geodesic (CNG) for co-rotating orbits on the equatorial plane of the Kerr background. We see that as we increase the values of $l=m$ the cloud approaches the value of the CNG in the limit $a/r_+ \rightarrow 1$. Recall that, as observed earlier, when $a/r_+ \rightarrow 1$, $r_0$ also tends  to the horizon. Thus, the position of these massless clouds can be arbitrarily close to the BH, in agreement with recent observations for the massive case~\cite{Benone:2014ssa,Hod:2014sha}.

\begin{widetext}

\begin{figure}[h!]
\centering
\includegraphics[height=2.4in]{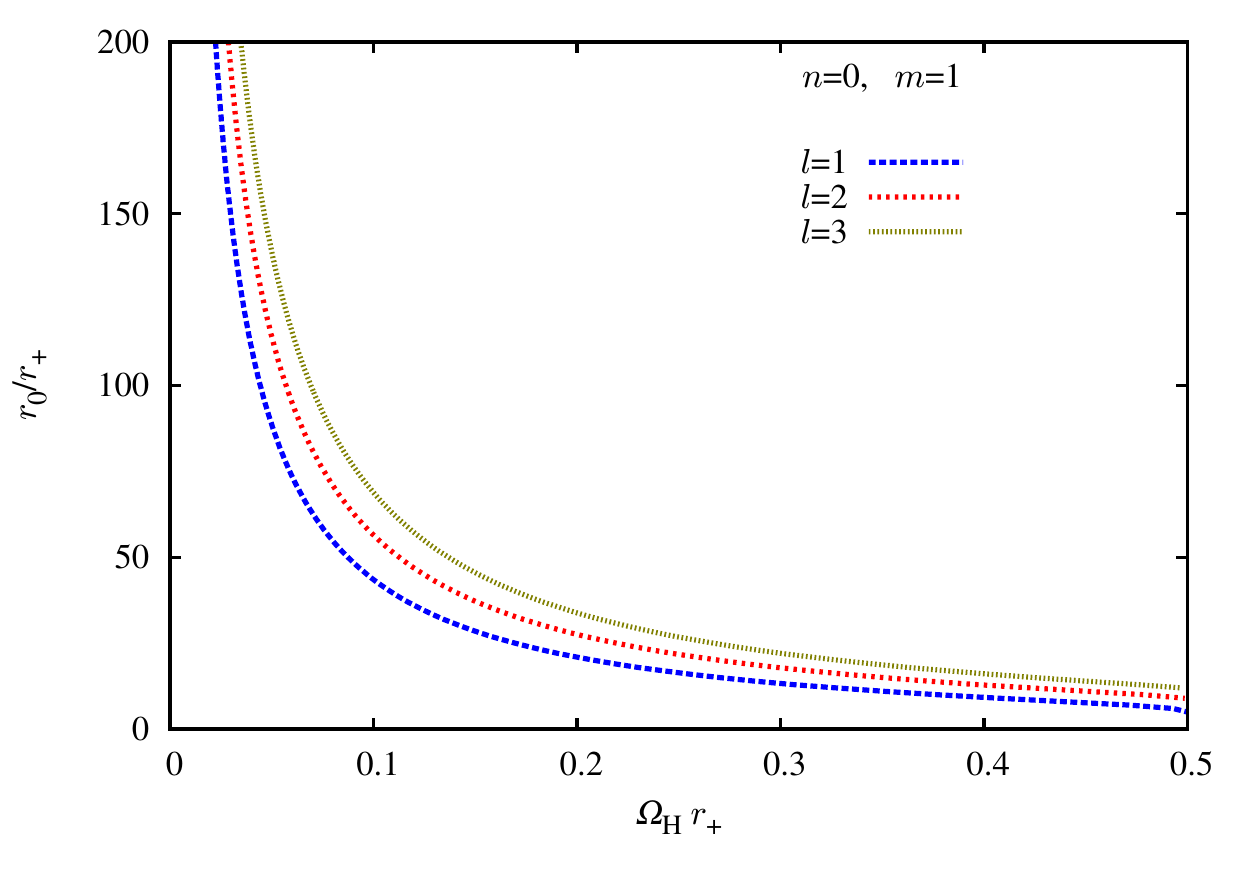}
\includegraphics[height=2.4in]{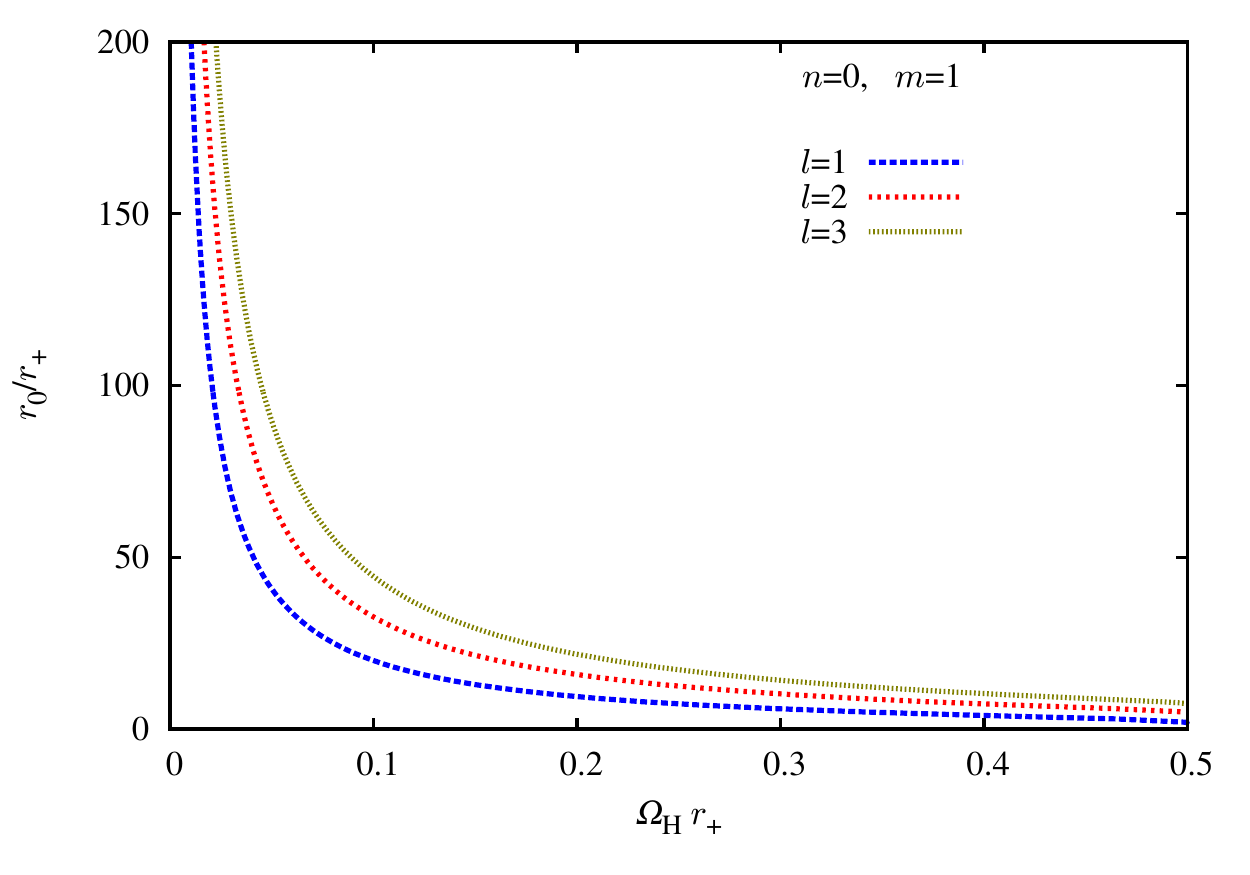} \\
\includegraphics[height=2.4in]{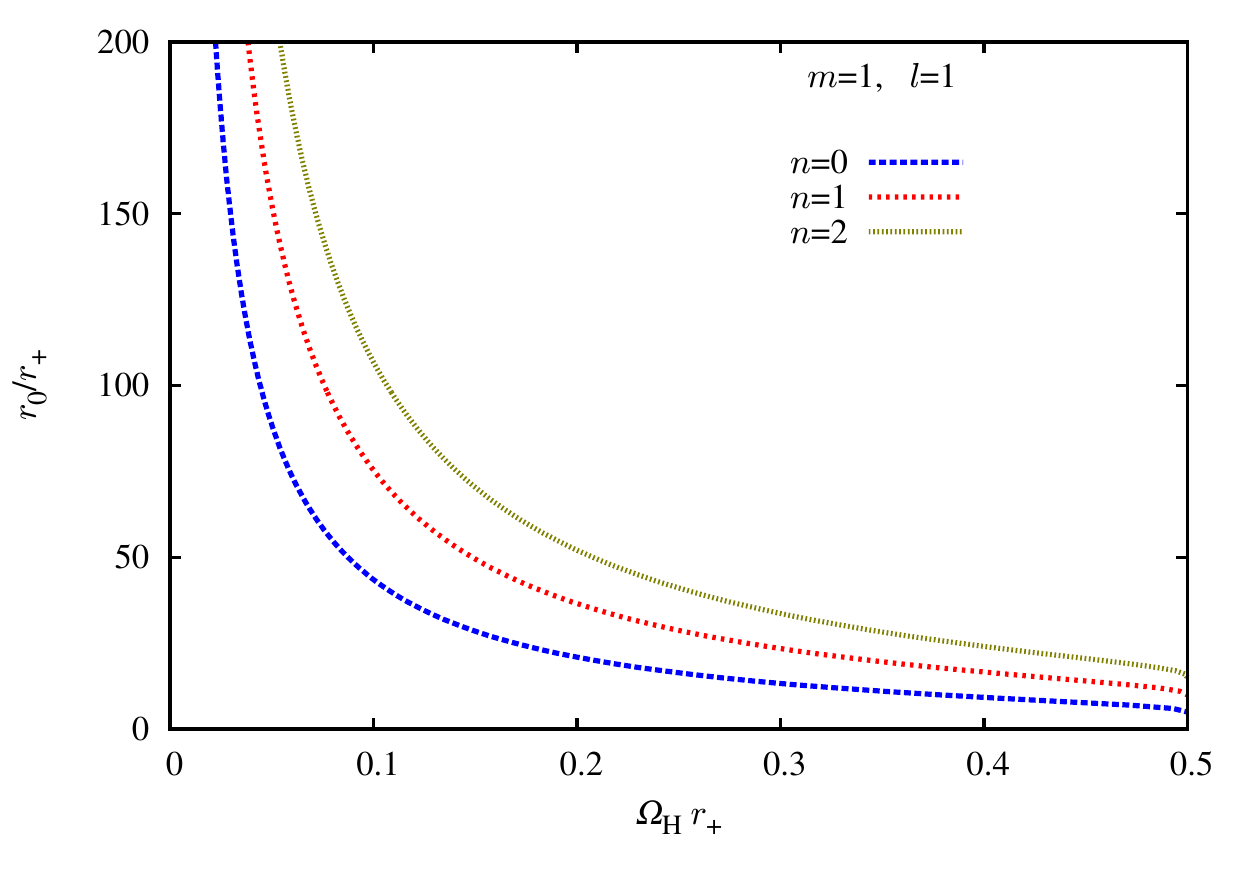}
\includegraphics[height=2.4in]{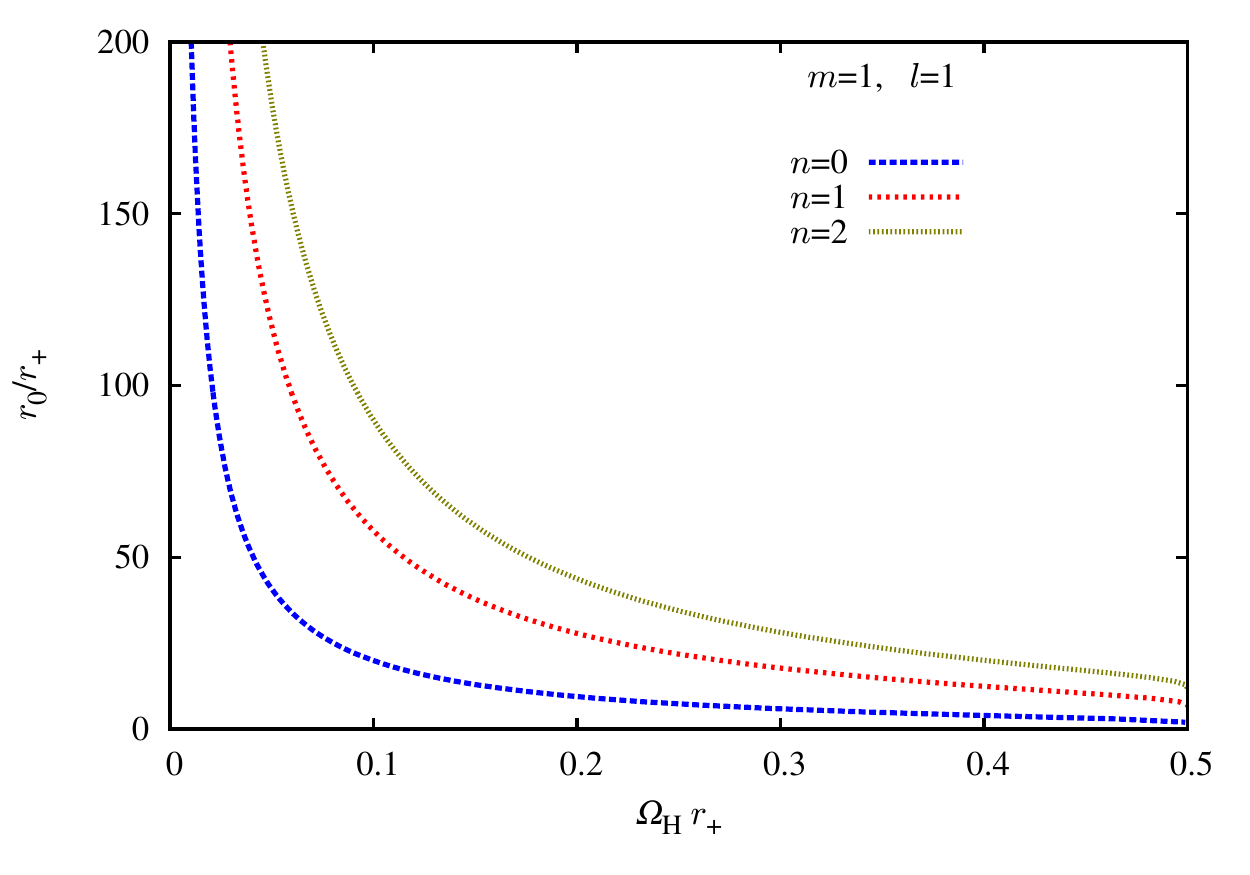} \\
\includegraphics[height=2.4in]{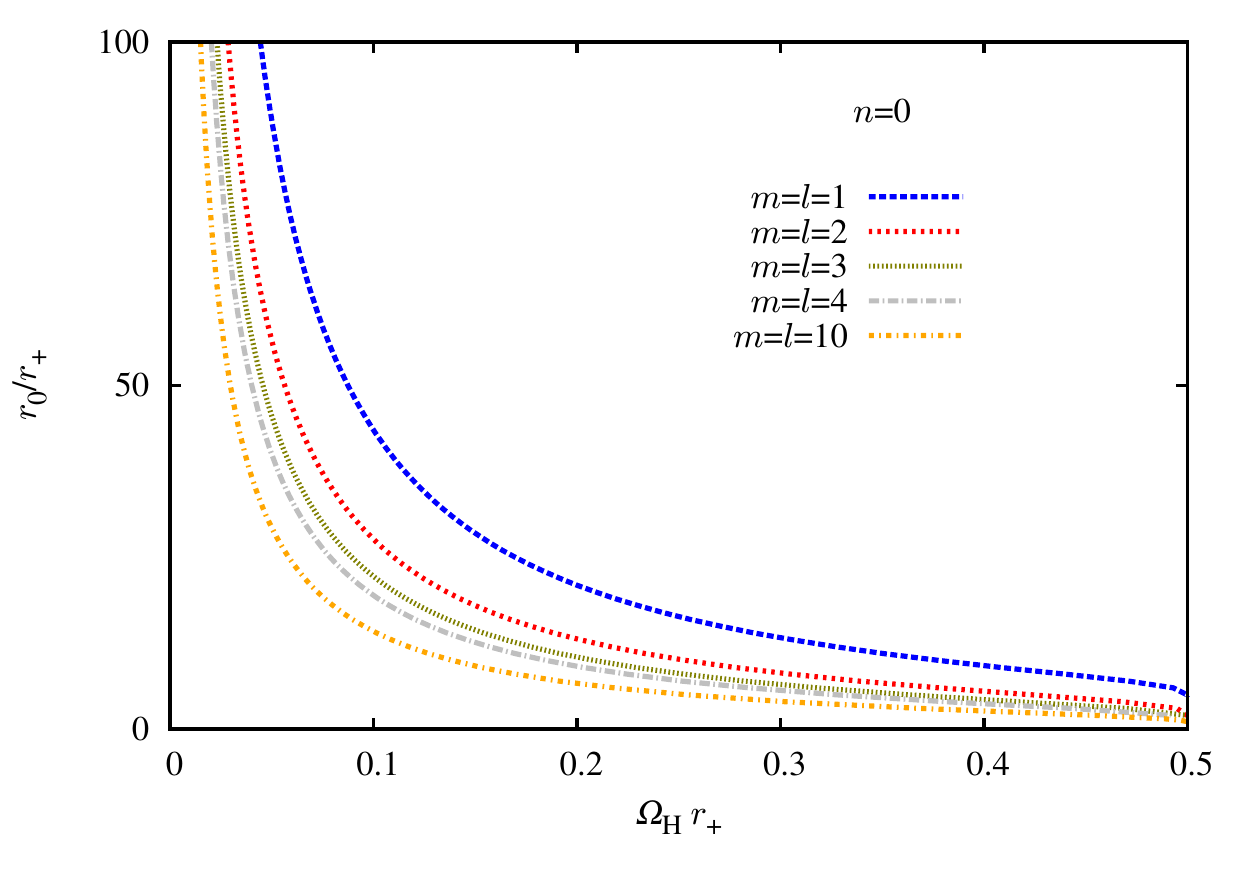}
\includegraphics[height=2.4in]{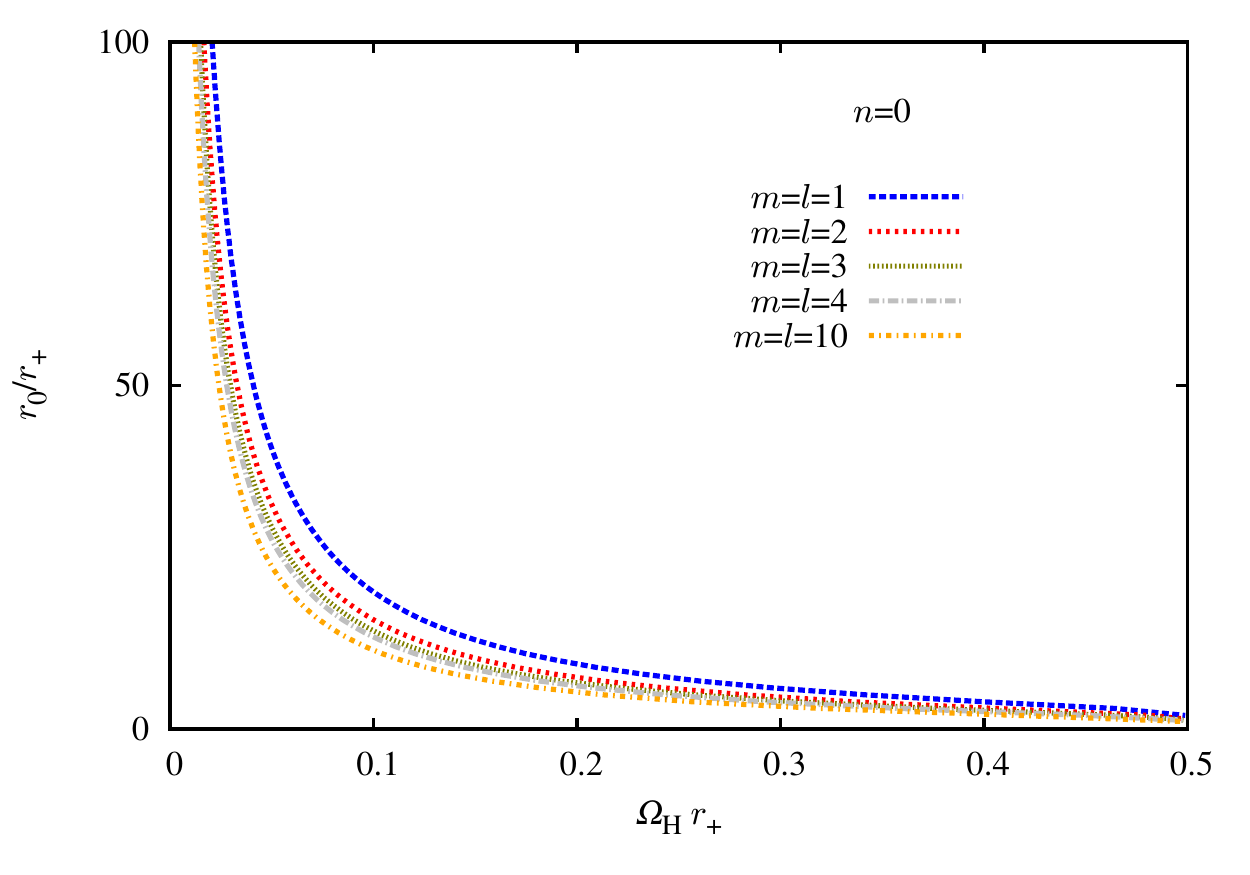}
\caption{Massless scalar clouds around Kerr BHs surrounded by a mirror at $r=r_0$ with a Dirichlet (left panels) or a Neumann (right panels) boundary condition. In each of the three panels we vary a subset of the cloud's `quantum numbers', as specified in the key.}
\label{fr0}
\end{figure}

\begin{figure}[h!]
\centering
\includegraphics[height=2.4in]{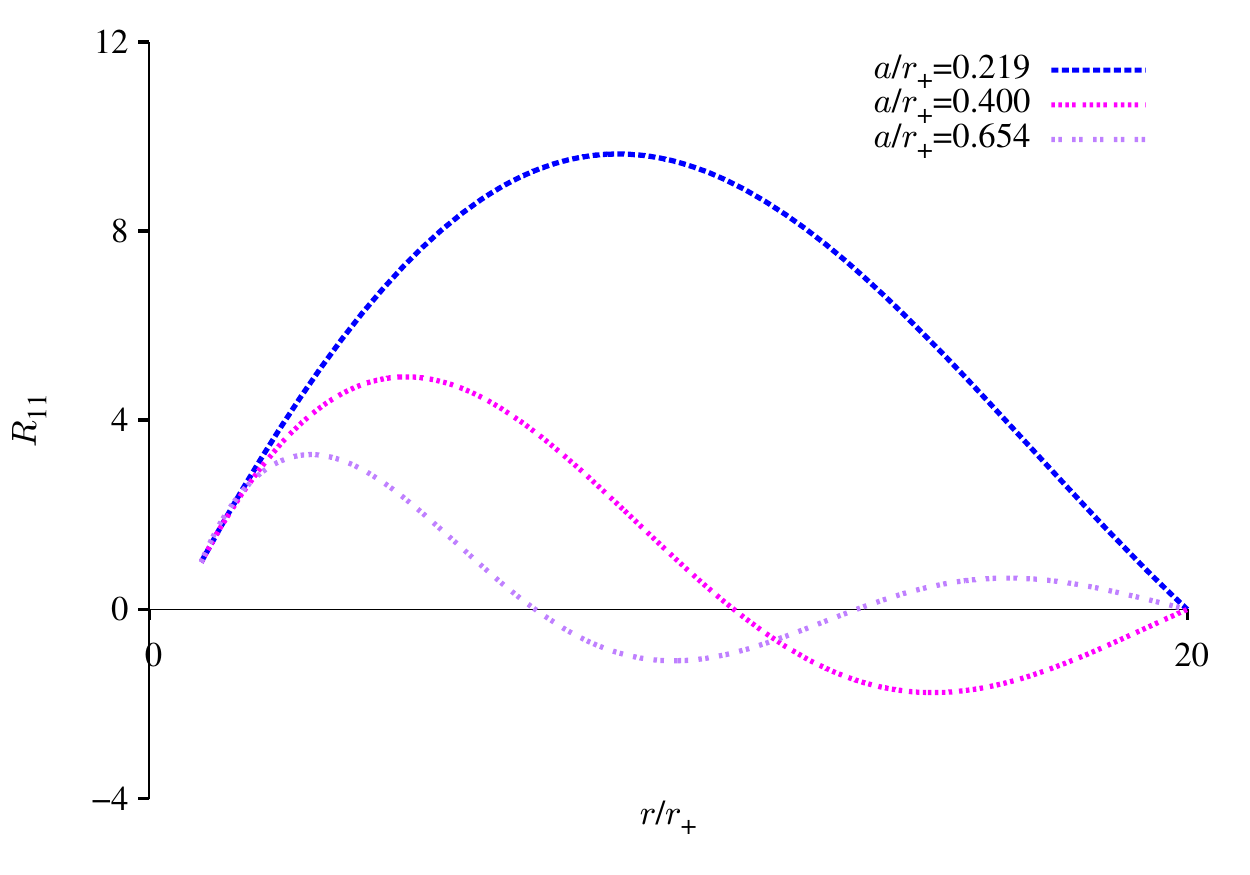}
\includegraphics[height=2.4in]{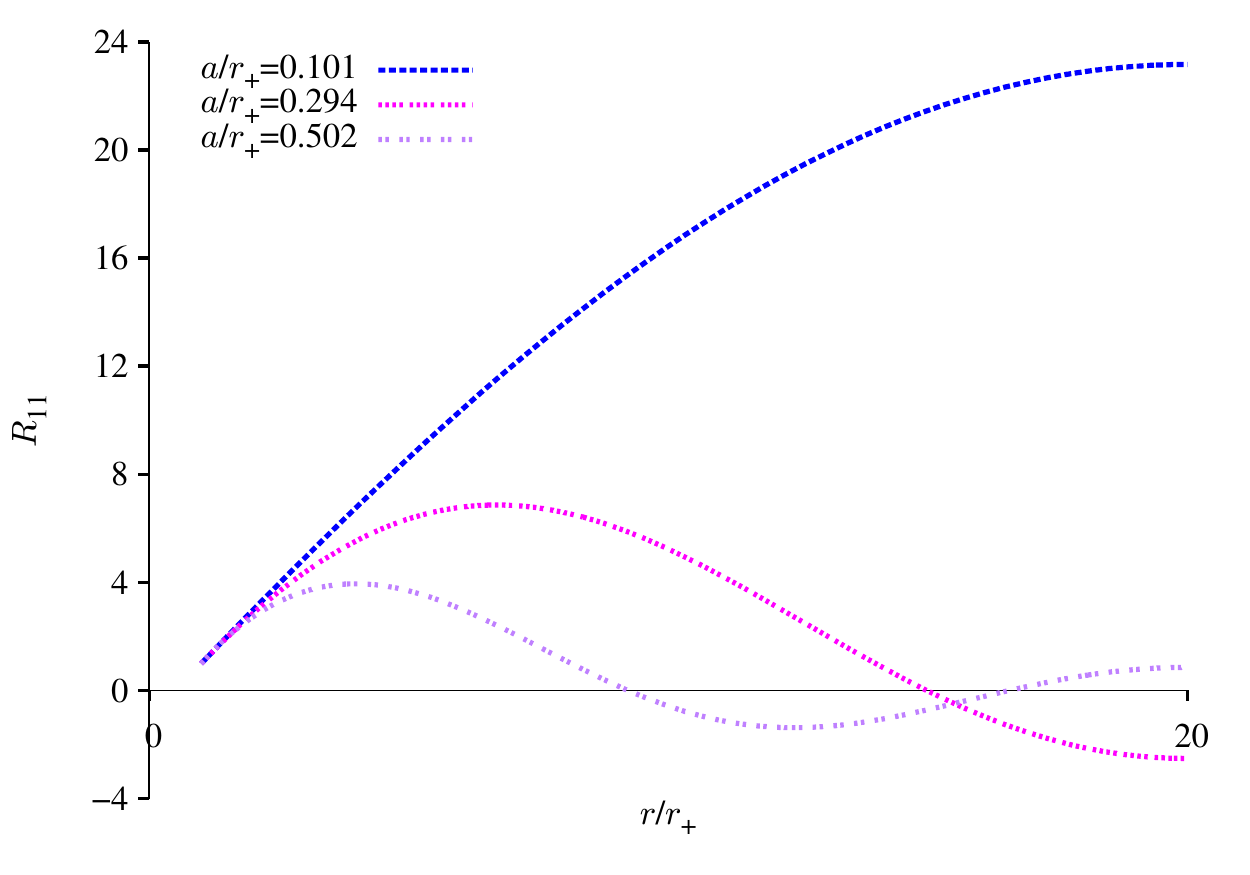} \\
\includegraphics[height=2.4in]{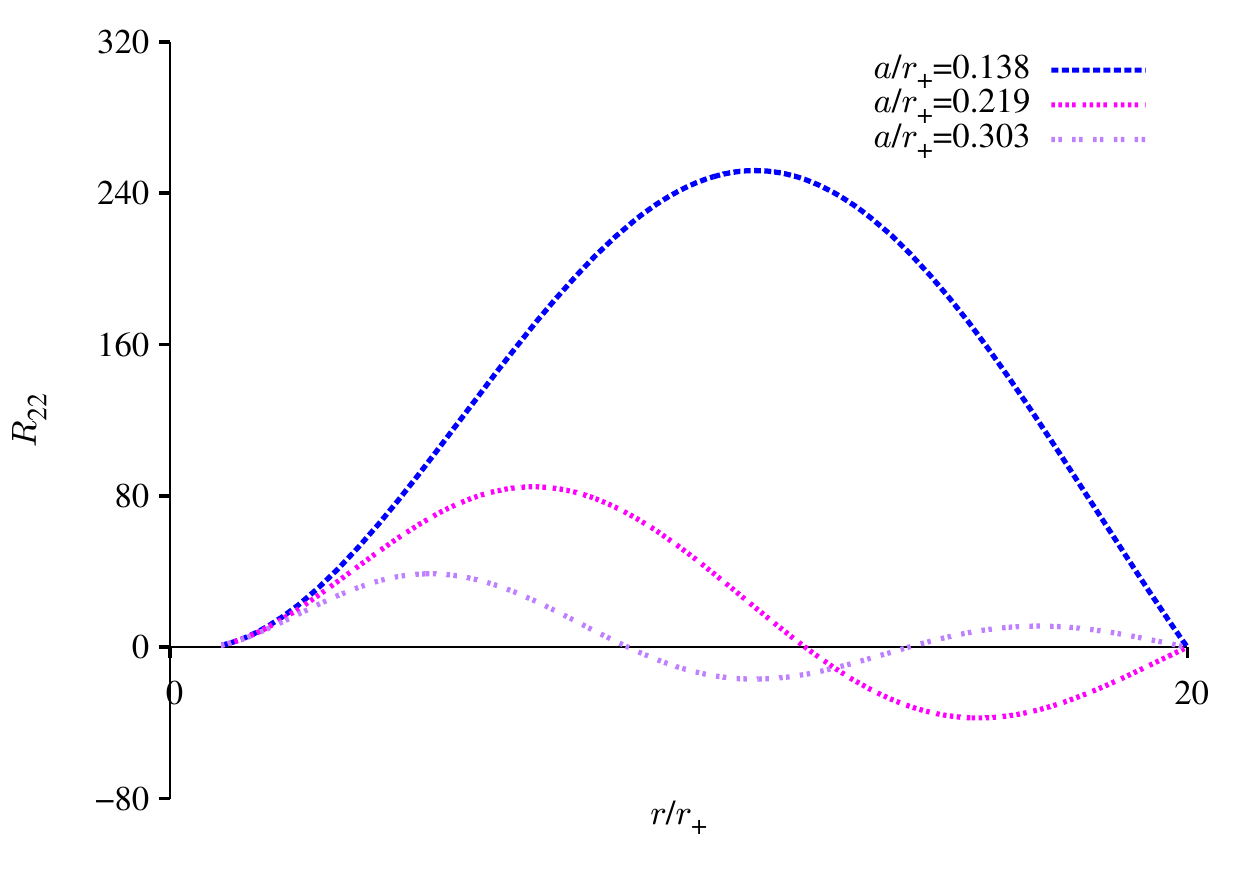}
\includegraphics[height=2.4in]{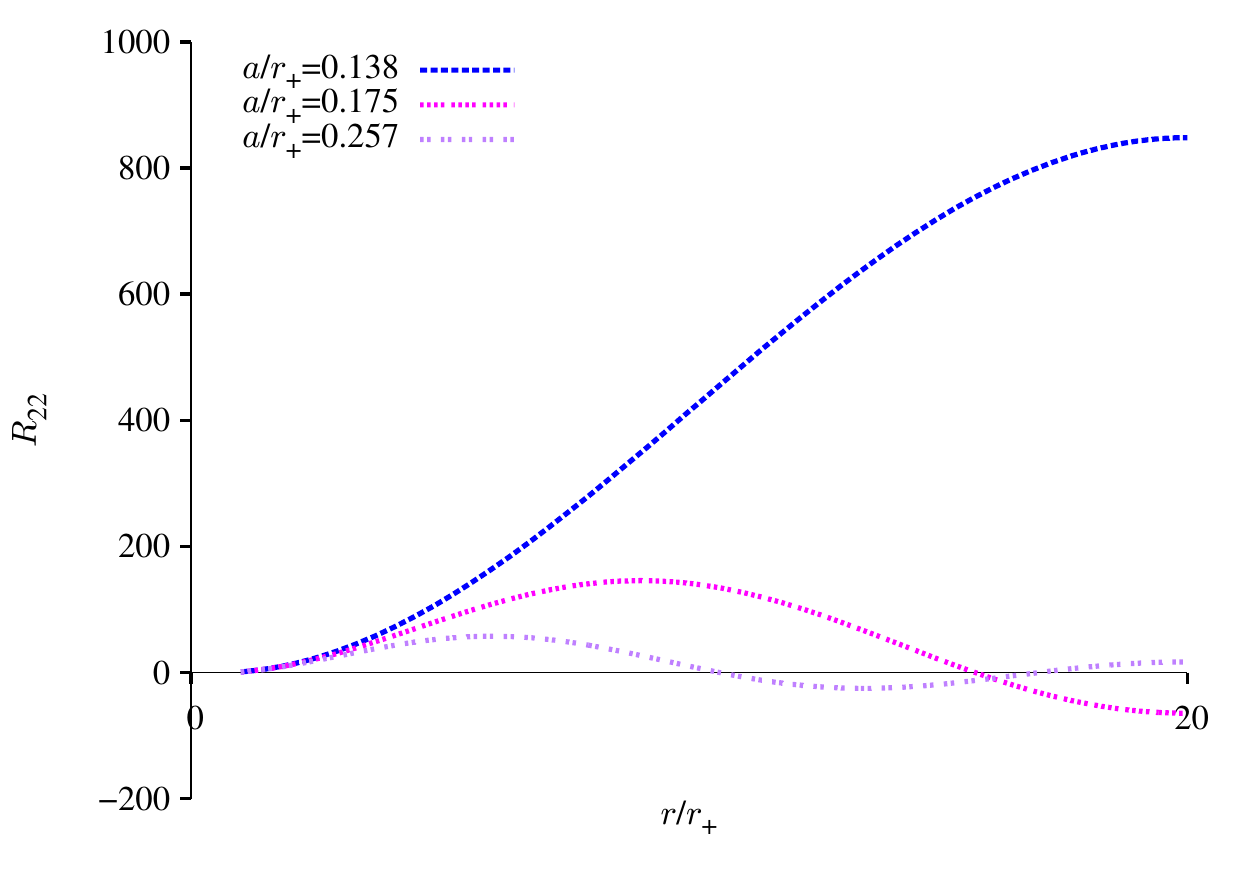}
\caption{Radial solutions $R_{11}$ and $R_{22}$ with $r_0/r_+=20$ for Dirichlet (left panels) and Neumann (right panels) boundary conditions.}
\label{frn}
\end{figure}

\begin{figure}[h!]
\centering
\includegraphics[height=2.4in]{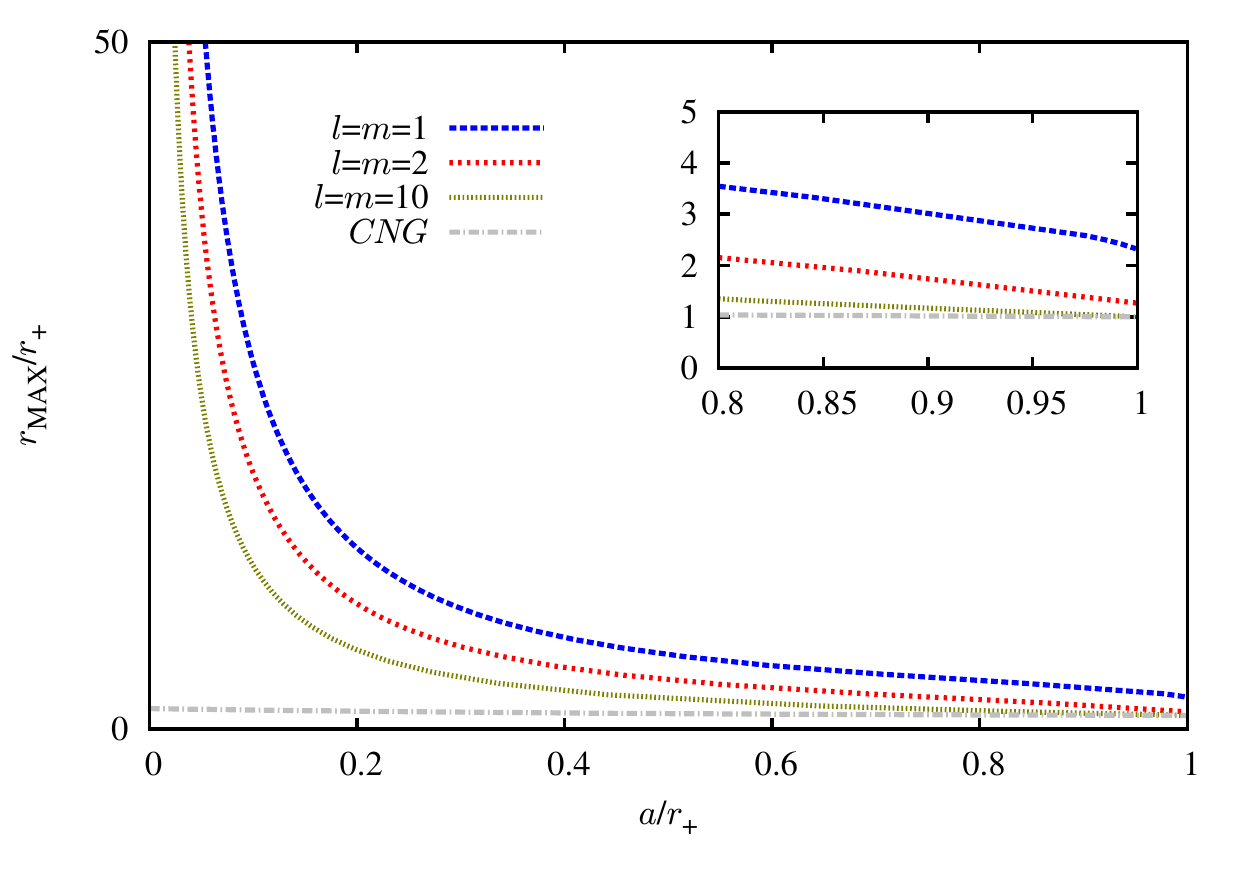}
\includegraphics[height=2.4in]{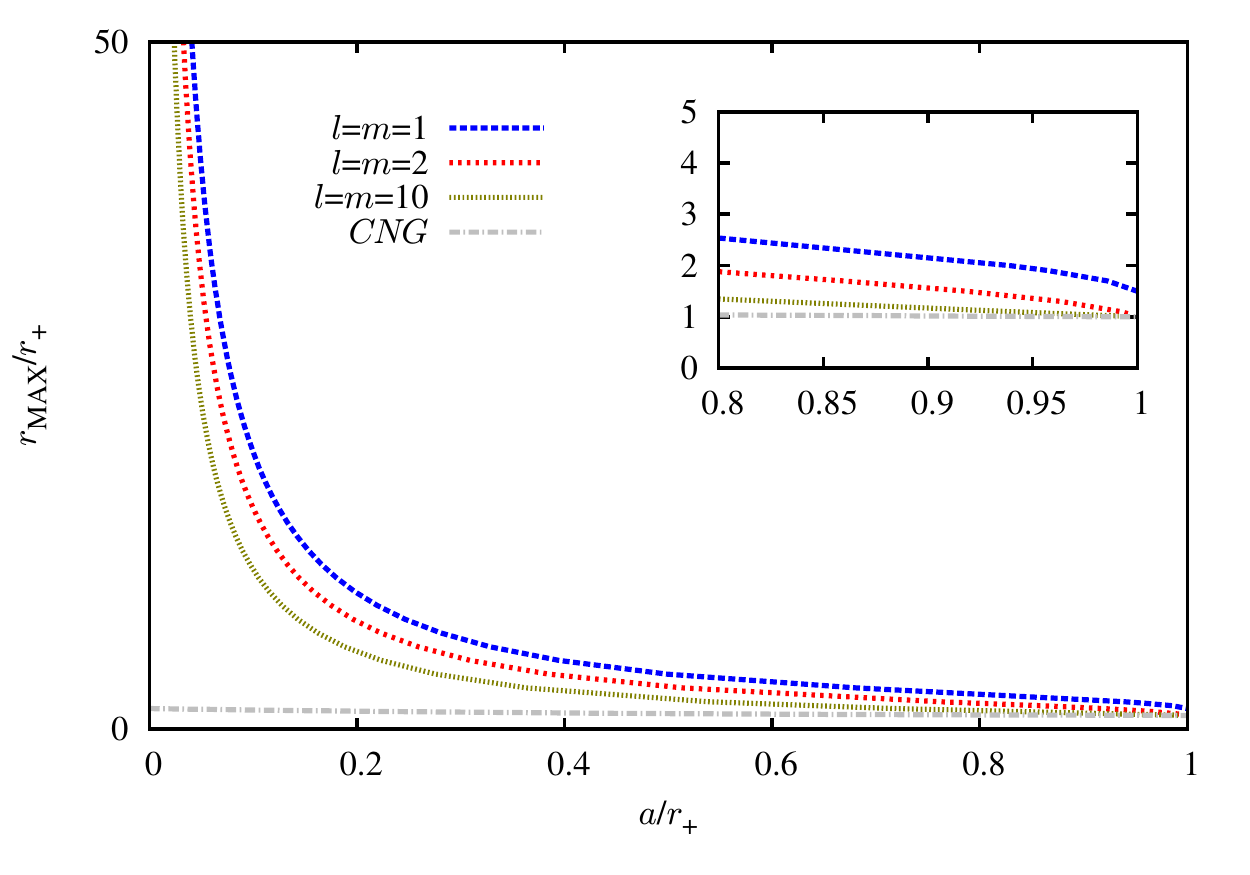}
\caption{Position of the clouds for different values of $l=m$ compared with the circular null geodesic (CNG) for Dirichlet (left panel) and Neumann (right panel) boundary conditions.}
\label{fok}
\end{figure}

\end{widetext}

Let us comment that we have verified qualitatively similar clouds exist not only for \textit{massive} but also for \textit{self-interacting} scalar fields around Kerr BHs in cavity, generalizing the ones studied in~\cite{Herdeiro:2014pka} for asymptotically flat spacetimes.

%%%%%%%%%%%%%%%%%%%%%%%
\section{Acoustic clouds in a draining bathtub}
\label{sec_acoustic}
%%%%%%%%%%%%%%%%%%%%%%%

\subsection{Background and acoustic perturbations}

We now turn to the acoustic BH analogue obtained from the the draining bathtub model considered in~\cite{Visser:1997ux}. The scalar field now describes acoustic perturbations on a fluid flow. We still consider it to be a complex scalar field, but all the results apply for a real scalar field as well, simply by taking the real (or imaginary) part of the result. The effective acoustic geometry corresponds to the metric seen by sound waves traveling in a fluid with flow velocity given by
\b
\vec{v} = \frac{A}{r}  \hat{r} + \frac{B}{r}  \hat{\phi}\ ,
\e
where the constants $A$ and $B$ are the draining and the circulation, respectively. For this model the following line element is obtained
\be
ds^2 = &-&\left(1-\frac{A^2+B^2}{r^2}\right)dt^2 + \left(1 - \frac{A^2}{r^2}\right)^{-1}dr^2\nonumber\\
 &-& 2B d\phi dt + r^2d\phi^2\ .
\label{dsa}
\ee
Thus, the effective geometry is 1+2 dimensional. It has an event horizon at $r_+=A$; this horizon is rotating with angular velocity $\Omega_H=B/A^2$.

In order to find the solution of the Klein-Gordon equation we again consider a Fourier/mode decomposition. The modes have the form
\b
\Phi_m(r,\phi,t) = e^{i(m \phi-\omega t)}\psi_m(r),
\e
where $\psi_m(r)$ obeys
\begin{widetext}
\b
\frac{1}{r}\left(1-\frac{A^2}{r^2}\right)\frac{d}{dr}\left[r\left(1-\frac{A^2}{r^2}\right)\frac{d\psi_m}{dr}\right] + \left[\omega^2-\frac{2Bm\omega}{r^2}-\frac{m^2}{r^2}\left(1-\frac{A^2+B^2}{r^2}\right)\right]\psi_m = 0.
\label{era}
\e
\end{widetext}
We can rewrite Eq. (\ref{era}) as
\b
\frac{d^2}{dr_*^2}\zeta_m + \left[\left(\omega - \frac{B m}{r^2}\right)^2 - V_m(r)\right]\zeta_m = 0\ ,
\label{era2}
\e
with
\b
V_m(r) = \left(1-\frac{A^2}{r^2}\right)\left[\frac{m^2-1/4}{r^2}+\frac{5 A^2}{4 r^2}\right] \ ,
\e 
where $\zeta_m = \sqrt{r} \psi_m$ and the Regge-Wheeler coordinate can be defined by
\b
\frac{d}{dr_*} =\left(1-\frac{A^2}{r^2}\right)\frac{d}{dr}\ .
\e 
We find that the solutions at the horizon and at infinity behave similarly as before
\b
\psi_{\omega m}(r) \approx 
\left\{ 
\begin{array}{ll}
e^{-i(\omega-\omega_c) r_*}, \quad &\mbox{for $r\rightarrow r_+$}\ ,\\
 e^{i\omega r_*}, \quad &\mbox{for $r\rightarrow \infty$}\ ,
\end{array}
\right.
\label{sol}
\e
with
\b
\omega_c = m B/A^2=m\Omega_H\ .
\e

%But again these solutions do not represent clouds. So we consider $\omega=\omega_c$ and $\psi(r_0) = 0$, %where $r_0$ is the position of the mirror. 

%Surface waves in a shallow basin also ``see" the effective metric (\ref{dsa}) \cite{Schutzhold:2002rf}. These waves travel with velocity $\sqrt{g h_\infty}$ where $g$ is the gravity acceleration and $h_\infty$ is the height of the fluid far away from the sink. Since their velocity depends only on the height of the fluid, we can adjust $h_\infty$ in order to form an event horizon for low velocities, avoiding that the liquid gets turbulent. This is a great advantage compared with the acoustic hole, which requires supersonic velocities of the flow to form a horizon. 

\subsection{Numerical procedure and results}

Superradiant scattering of acoustic waves can be seen in this type of geometries, without the need of introducing a cavity~\cite{Crispino:2007zz,BCL,Basak:2002aw}. To obtain acoustic clouds, however, we need to consider $\omega=\omega_c$ and introduce a mirror at $r=r_0$ where boundary conditions of Dirichlet ($\psi_m(r_0) = 0$) or Neumann [$d(\psi_m(r))/dr$] types are imposed.\footnote{These boundary conditions are associated to a cylindrical mirror with low (Dirichlet) or high (Neumann) acoustic impedance \cite{Lax1948,Oliveira:2014oja}, the latter being perhaps more easily implemented experimentally, since it corresponds to a rigid boundary cylinder.} In the present case the clouds will be labelled by only two `quantum numbers': the azimuthal harmonic index $m$ and the node number $n$.

The numerical procedure is analogous to that adopted in Subsec.~\ref{nprkerr}.  We consider the power series expansion 
\b
\label{rah2}
%&\ & R_{lm}=R_0\bigg(1+\sum_{k\geq 1}R_k(r-r_+)^k\bigg),\nonumber\\
 \psi_{m}=\psi_0\bigg(1+\sum_{k\geq 1}\psi_k(r-r_+)^k\bigg)
\e
close to the horizon. To find the coefficients $\psi_k$ we substitute Eq.~(\ref{rah2}) in Eq.~(\ref{era}) and expand it in terms of $(r-r_+)$, taking $\psi_0=1$. We end up with a shooting problem for the radial function with shooting parameter $B$. Fixing the quantum numbers ($n$ and $m$) and using the event horizon $r_+=A$ as a scaling factor we choose values for $r_0$, solve the radial equations from close to the event horizon until the position of the mirror and find the values of $B$ that obey the appropriate boundary condition. 

In Fig. \ref{fra} we compare clouds with different quantum numbers for Dirichlet and Neumann boundary conditions. The observed behaviour is completely analogous to that seen for the Kerr case discussed above: (i) as $r_0/r_+ \rightarrow \infty$, $\Omega_H \rightarrow 0$; (ii) as we approach the mirror to the acoustic hole $\Omega_H$ increases; (iii) as $n$ is increased or $m$ is decreases, the angular velocity $\Omega_H A$ increases for fixed mirror position $r_0/A$; (iv) for the Neumann boundary condition the existence lines occur for smaller values of $\Omega_H$, for the same $r_0$.

In Fig. \ref{acoustic_profile1} and Fig. \ref{acoustic_profile2} we exhibit 3D plots with the radial and azimuthal profile of some examples of acoustic clouds for the Dirichlet boundary condition, without and with nodes, respectively. In Fig. \ref{acoustic_profile3} and Fig. \ref{acoustic_profile4} we exhibit the same as in Fig. \ref{acoustic_profile1} and Fig. \ref{acoustic_profile2}, but for the Neumann boundary condition. 

One may wonder if an analytic solution may be found for Eq. (\ref{era}). Studies of the draining bathtub model have actually concluded that the radial function in this case can be expressed in terms of Heun functions (see $e.g.$~\cite{Vieira:2014rva}). It seems, however, that not much use can be made of these functions to find the relevant information to obtain the acoustic clouds. Moreover, for the acoustic hole, there is no theoretical restriction that prevents taking $B>A$; in other words, there is no extremal case for this analogue model, for which there could be some extra simplification for the radial equation, as it happens for the clouds around extremal Kerr BHs~\cite{Hod:2012px}.

\begin{widetext}

\begin{figure}[h!]
\centering
\includegraphics[height=2.4in]{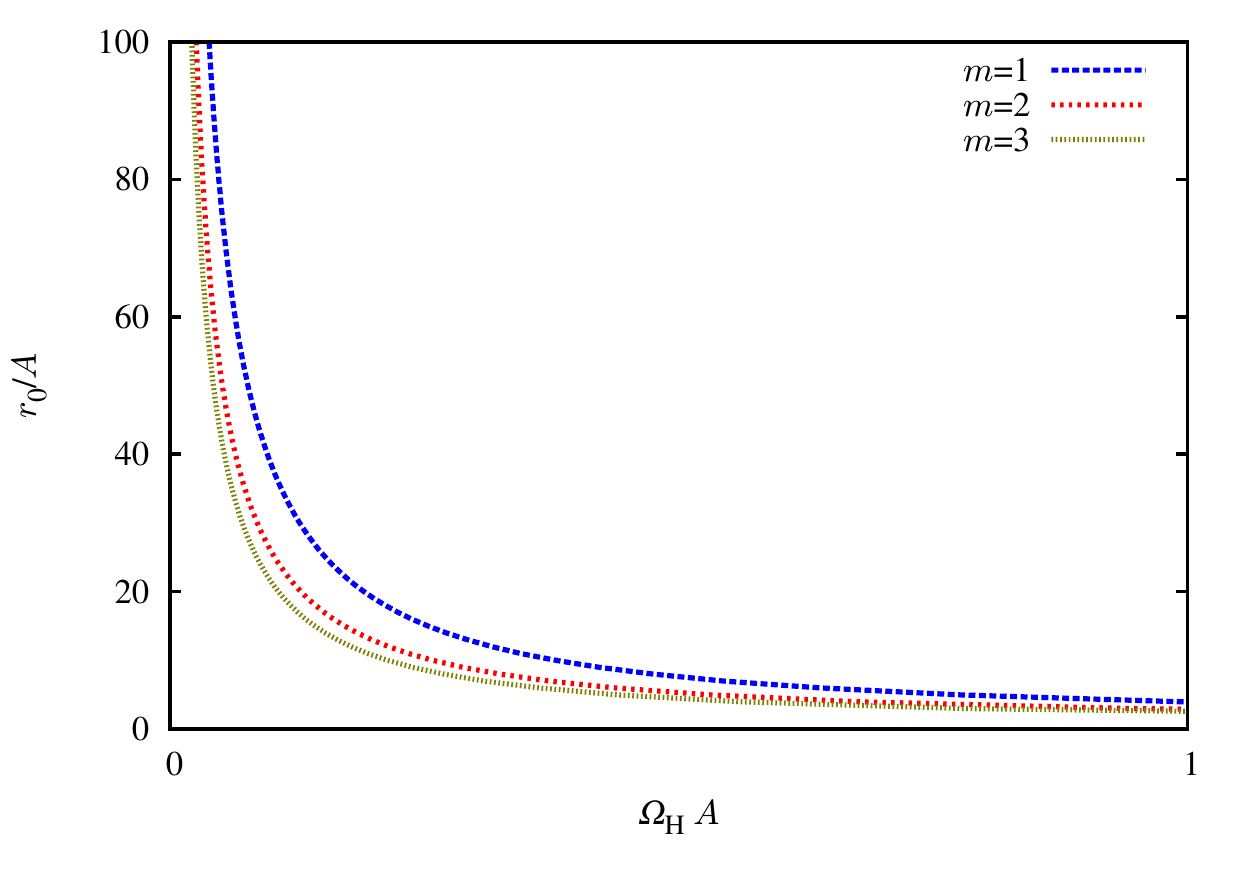}
\includegraphics[height=2.4in]{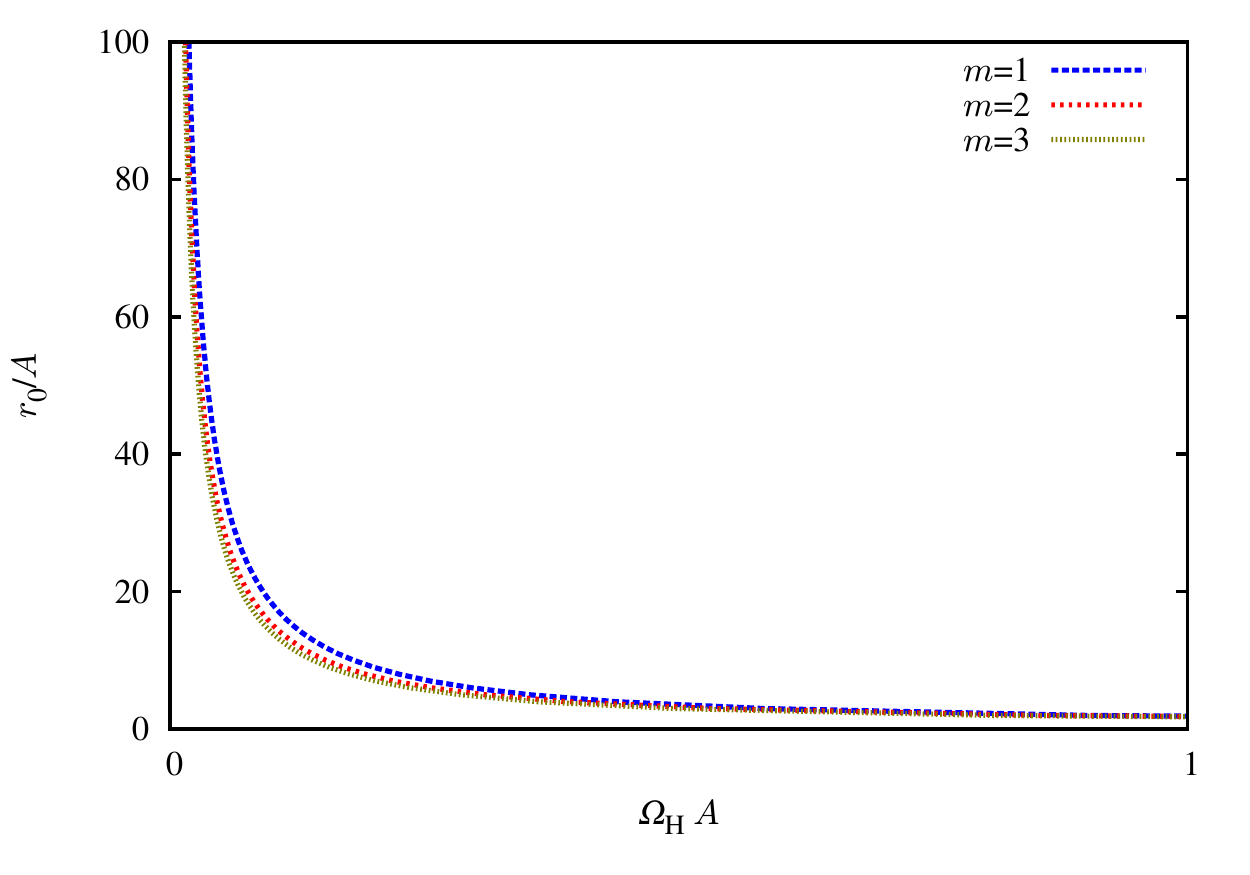} \\
\includegraphics[height=2.4in]{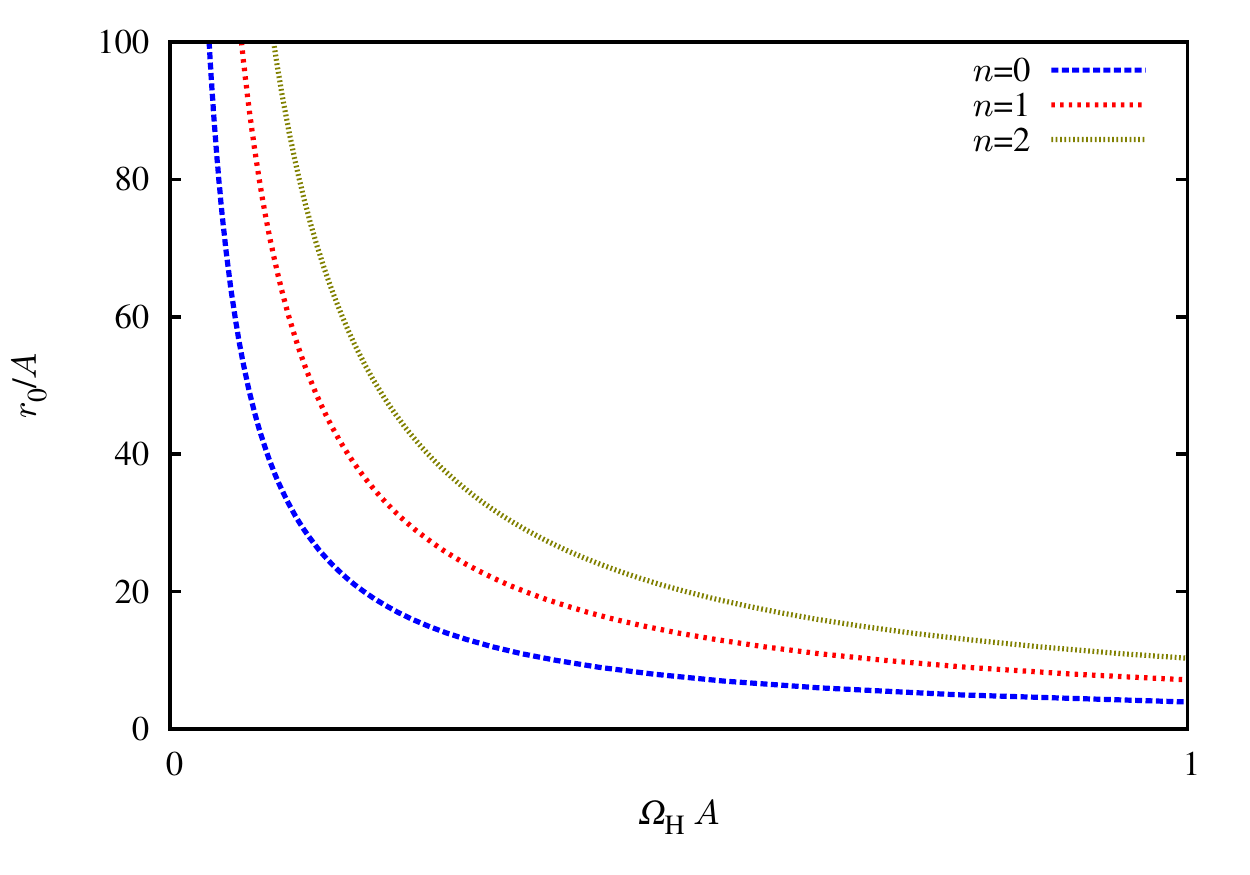}
\includegraphics[height=2.4in]{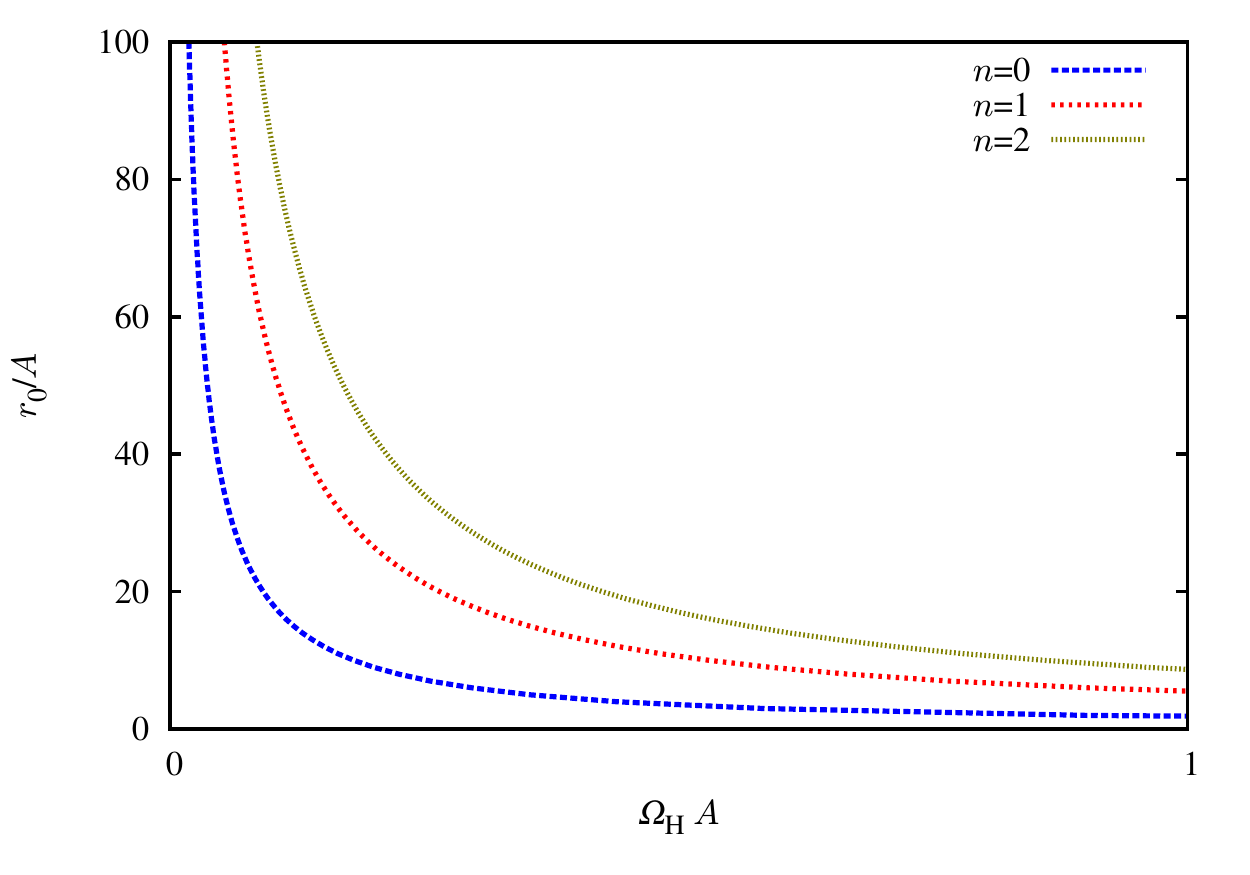}
\caption{Acoustic clouds on the draining bathtub surrounded by a mirror at $r=r_0$ for  Dirichlet  (left panels) boundary conditions and Neumann (right panels) boundary conditions. In each of the two panels we vary one of the two cloud's `quantum numbers', as specified in the key.}
\label{fra}
\end{figure}

%\begin{figure}[h!]
%\centering
%\includegraphics[height=2.5in]{bagnm.pdf}
%\includegraphics[height=2.5in]{bagnn.pdf}
%\caption{Same as in Fig. (\ref{fra}), but for the Newmann boundary conditions.} 
%\label{fran}
%\end{figure}

\begin{figure}[h!]
\centering
\includegraphics[height=2.6in]{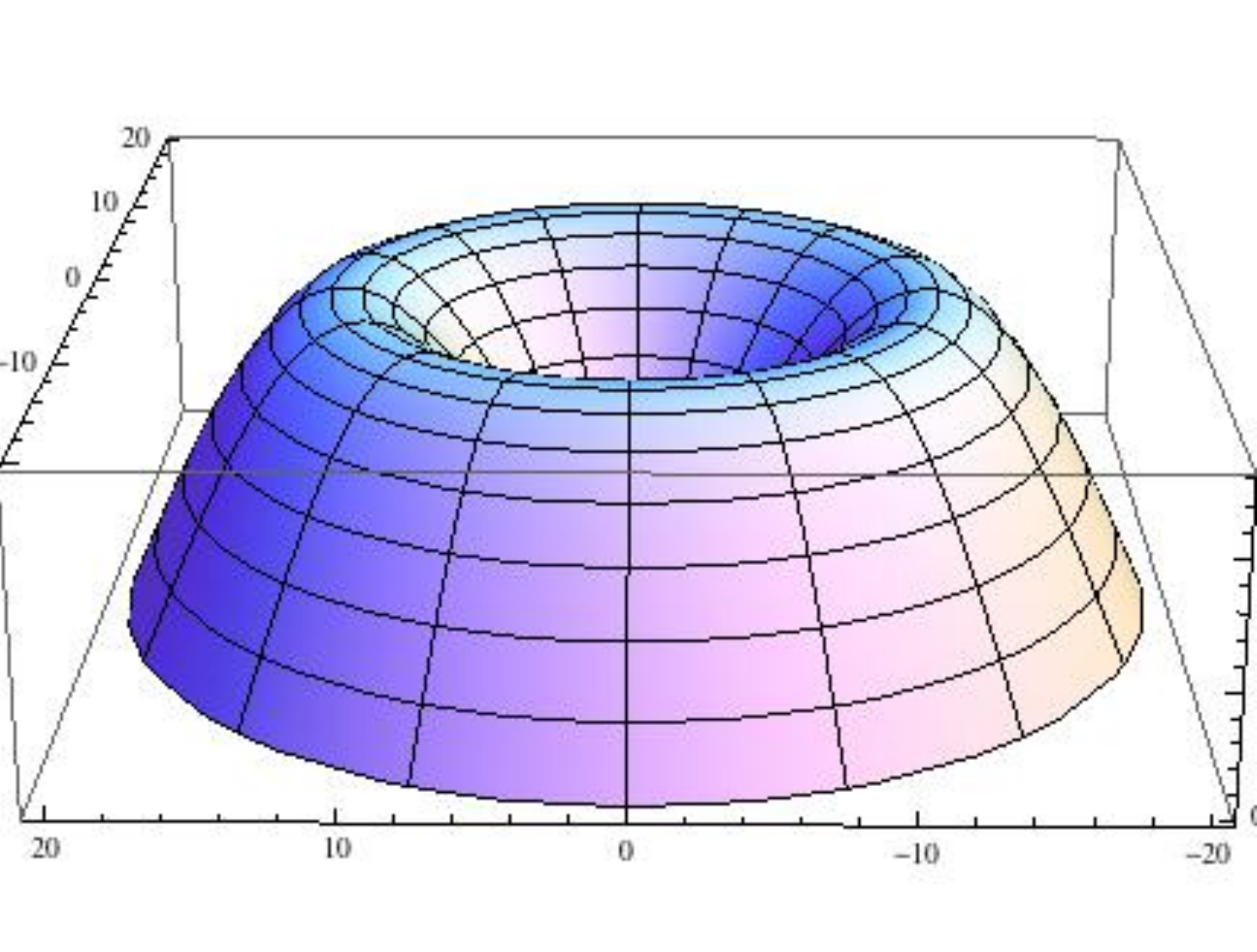}
\includegraphics[height=2.45in]{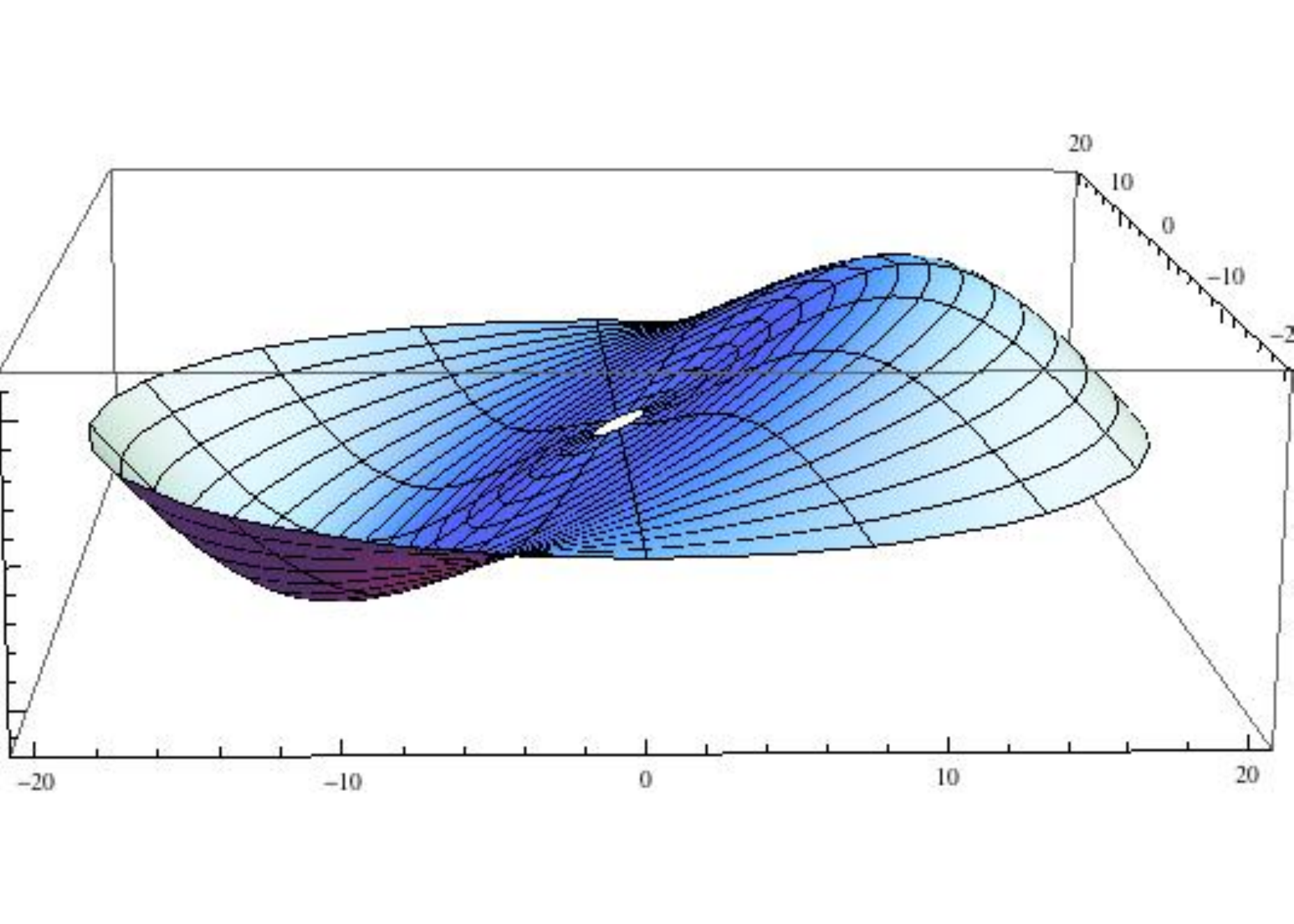} \\ 
\includegraphics[height=2.6in]{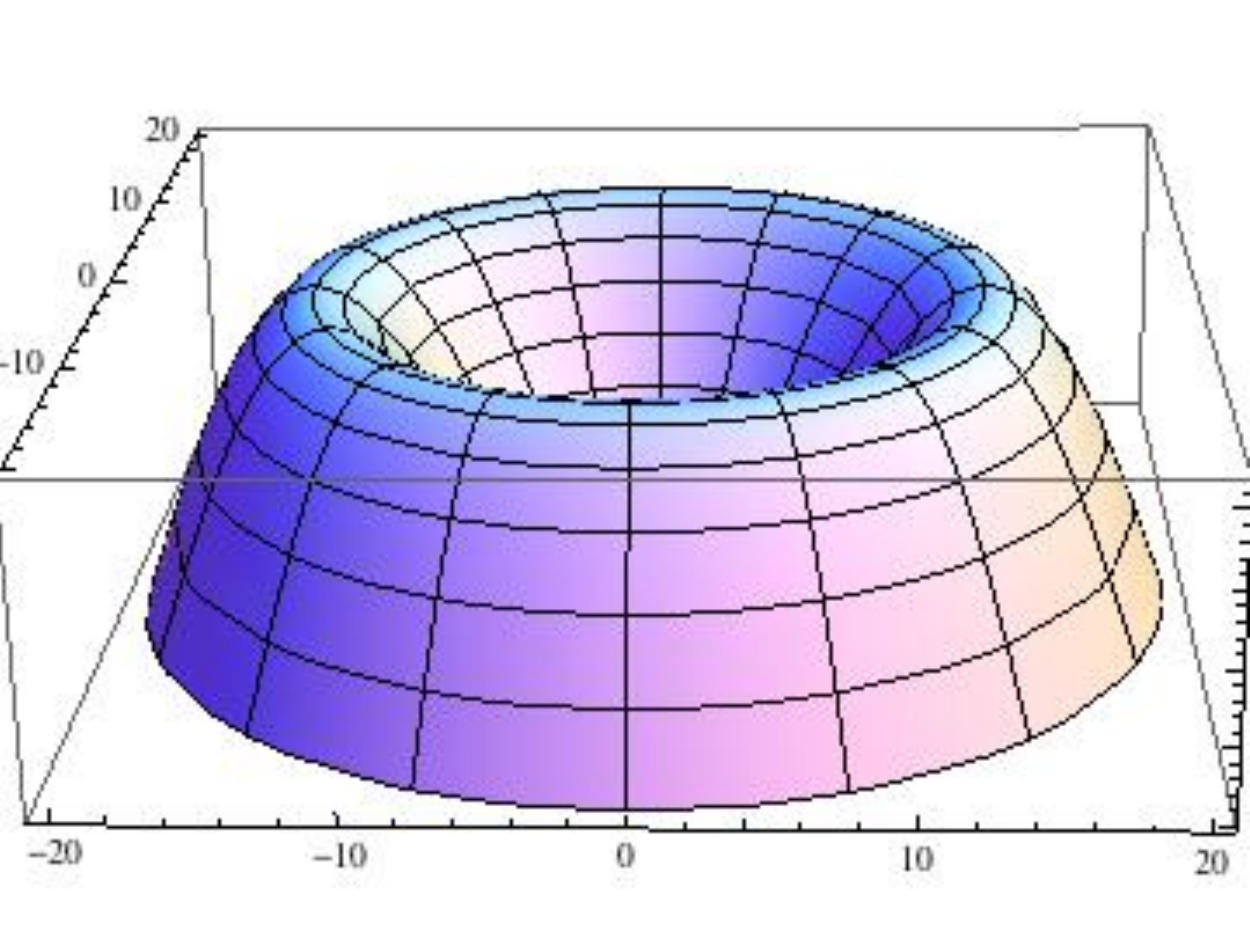}
\includegraphics[height=2.45in]{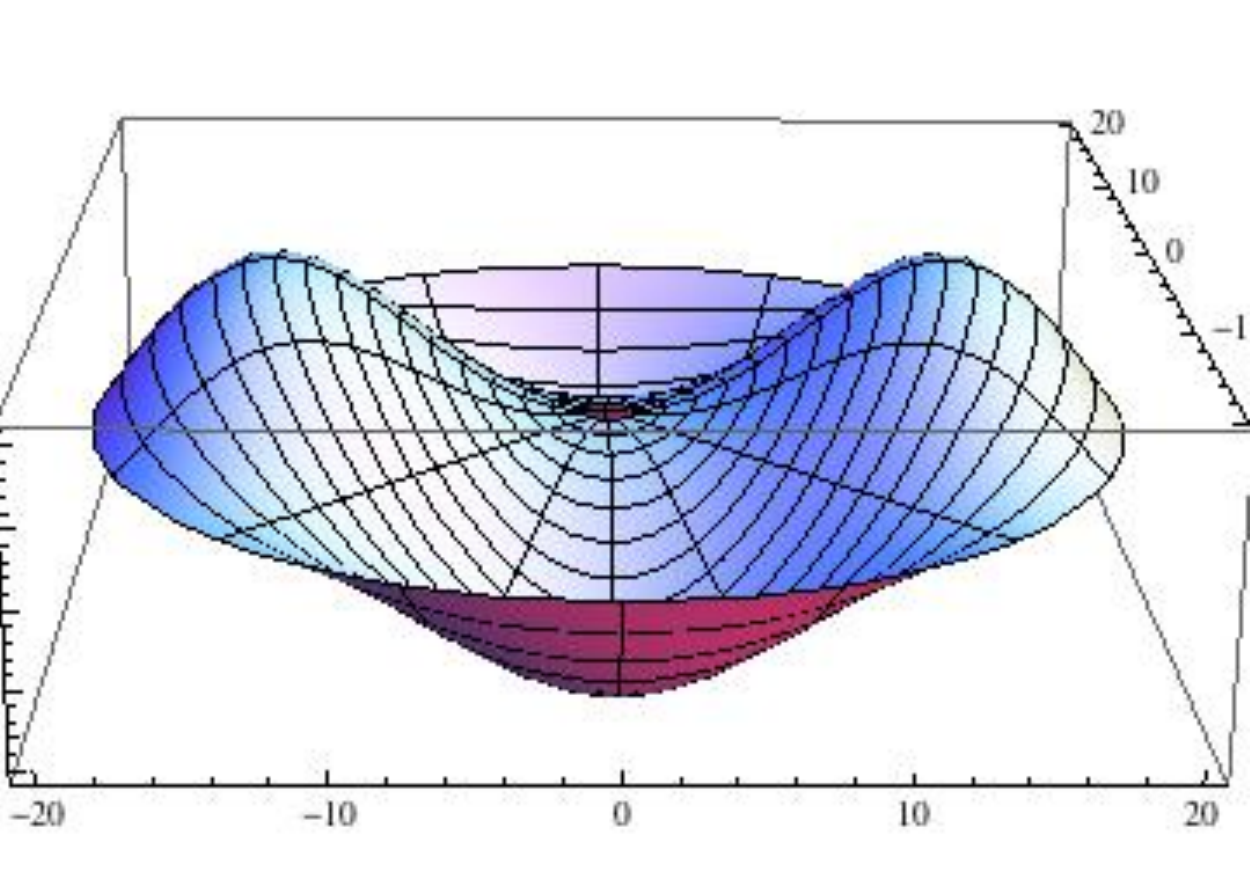} \\ 
\caption{Spatial distribution of nodeless acoustic clouds with Dirichlet boundary condition and $m=1$  (top panels)  or  $m=2$ (bottom panels), on the $r,\phi$ plane. On the left panels we present $|\Phi_m|$, while on the right panels we show the real part of $\Phi_m$ for a fixed $t$. In all plots the outer edge is the cylindrical mirror and the inner boundary is the acoustic BH horizon (clearly seen on the top right panel). We have fixed $r_0/A=20$ and the cloud supporting background has $B\simeq 0.19$ for the top panels and $B\simeq 0.13$ for the bottom ones.}
\label{acoustic_profile1}
\end{figure}

\begin{figure}[h!]
\centering
\includegraphics[height=2.6in]{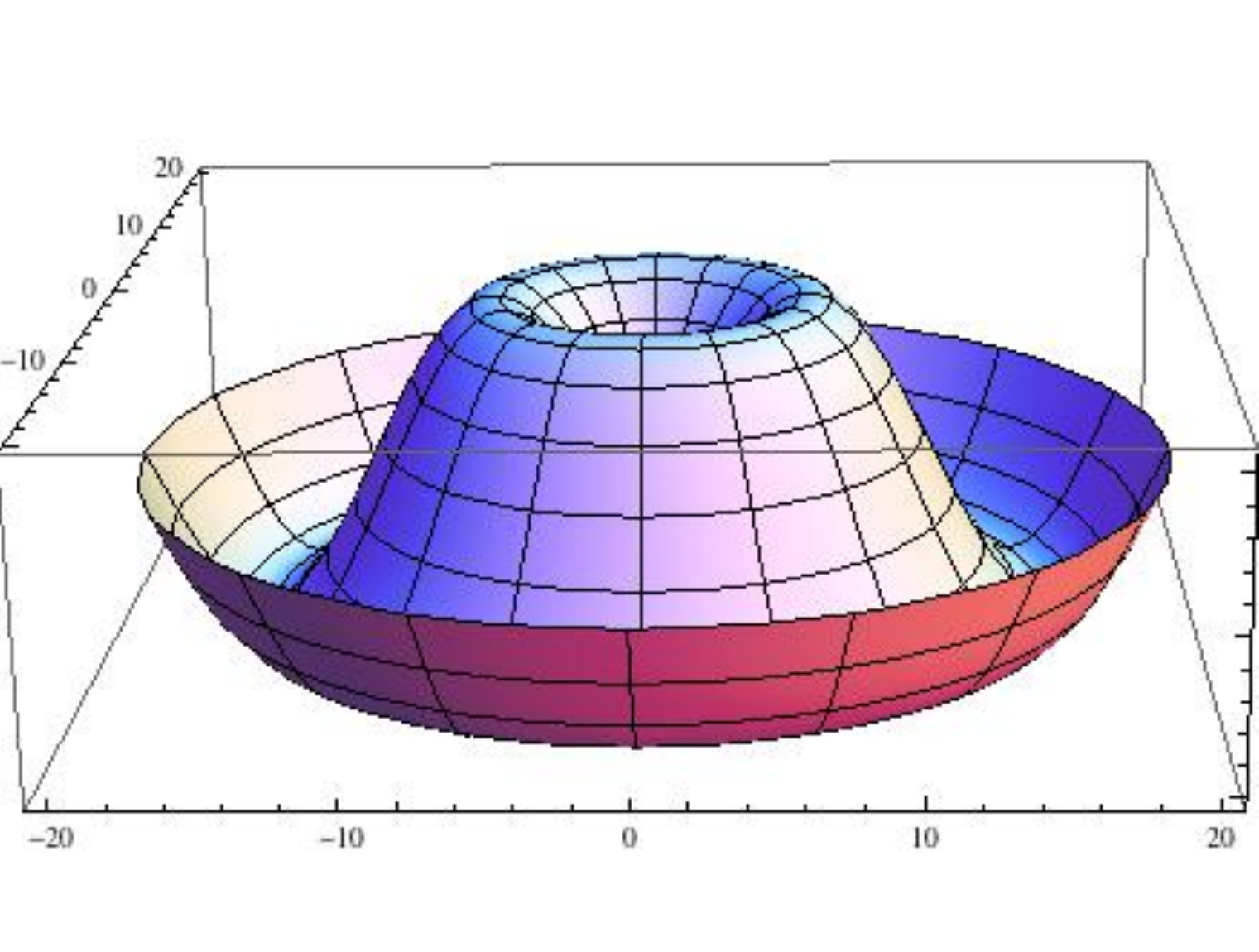}
\includegraphics[height=2.45in]{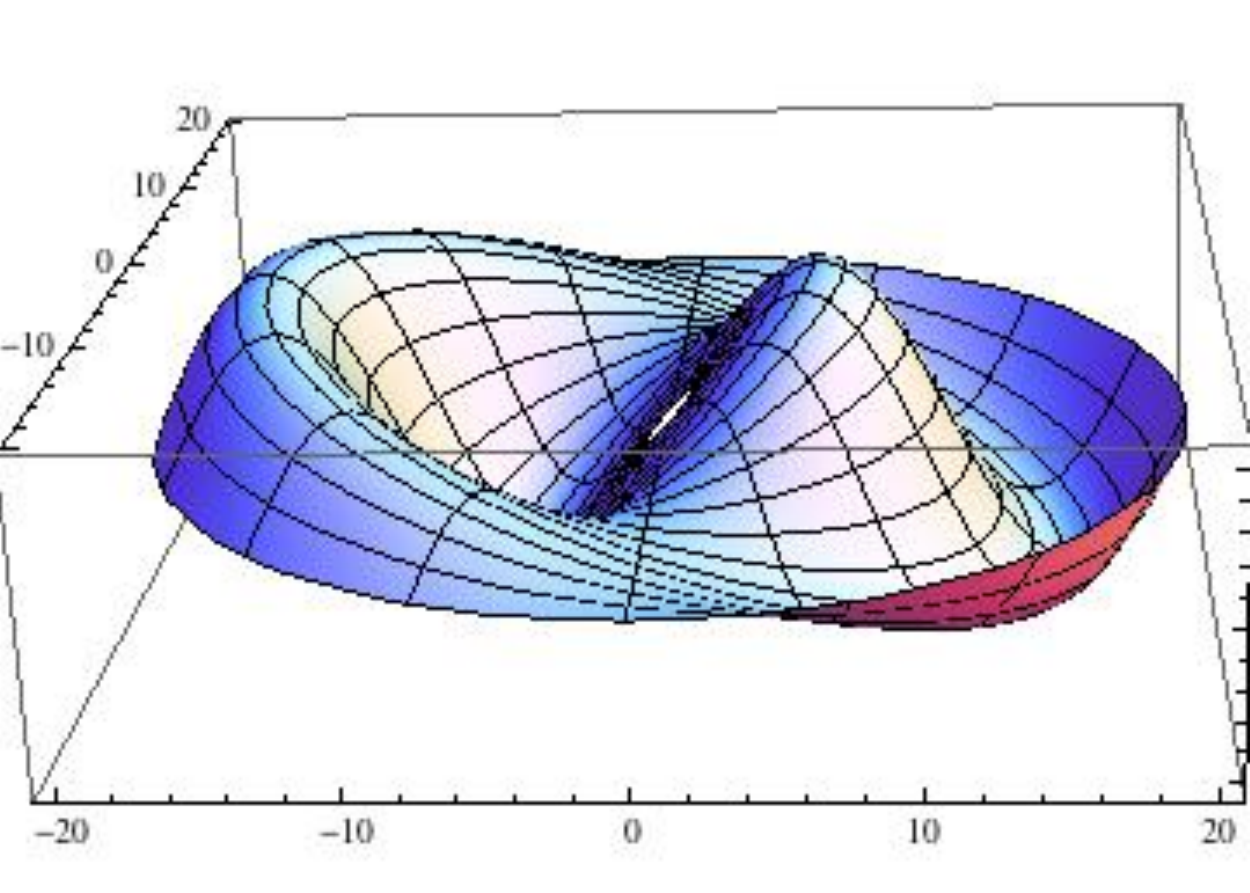} \\ 
\includegraphics[height=2.6in]{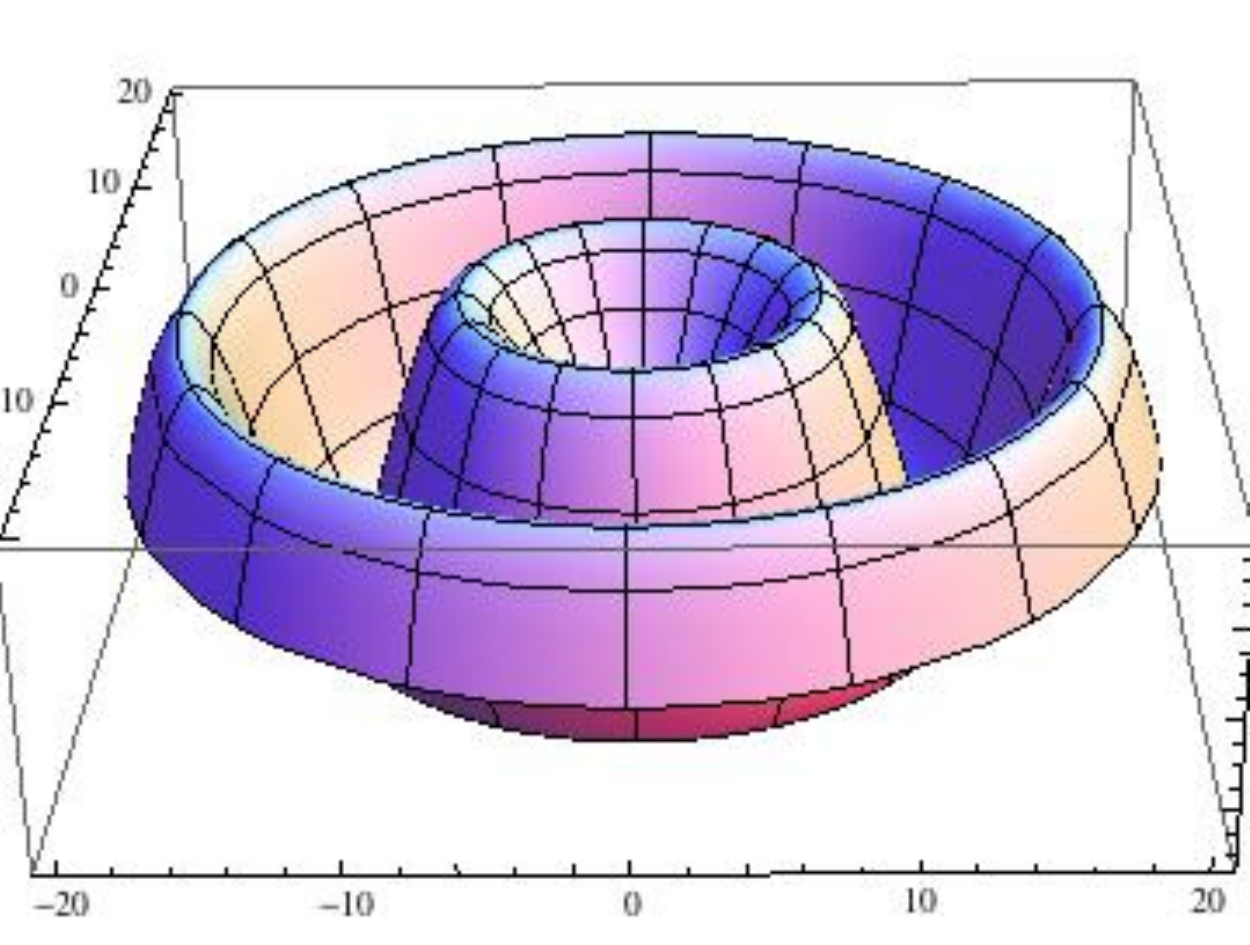}
\includegraphics[height=2.45in]{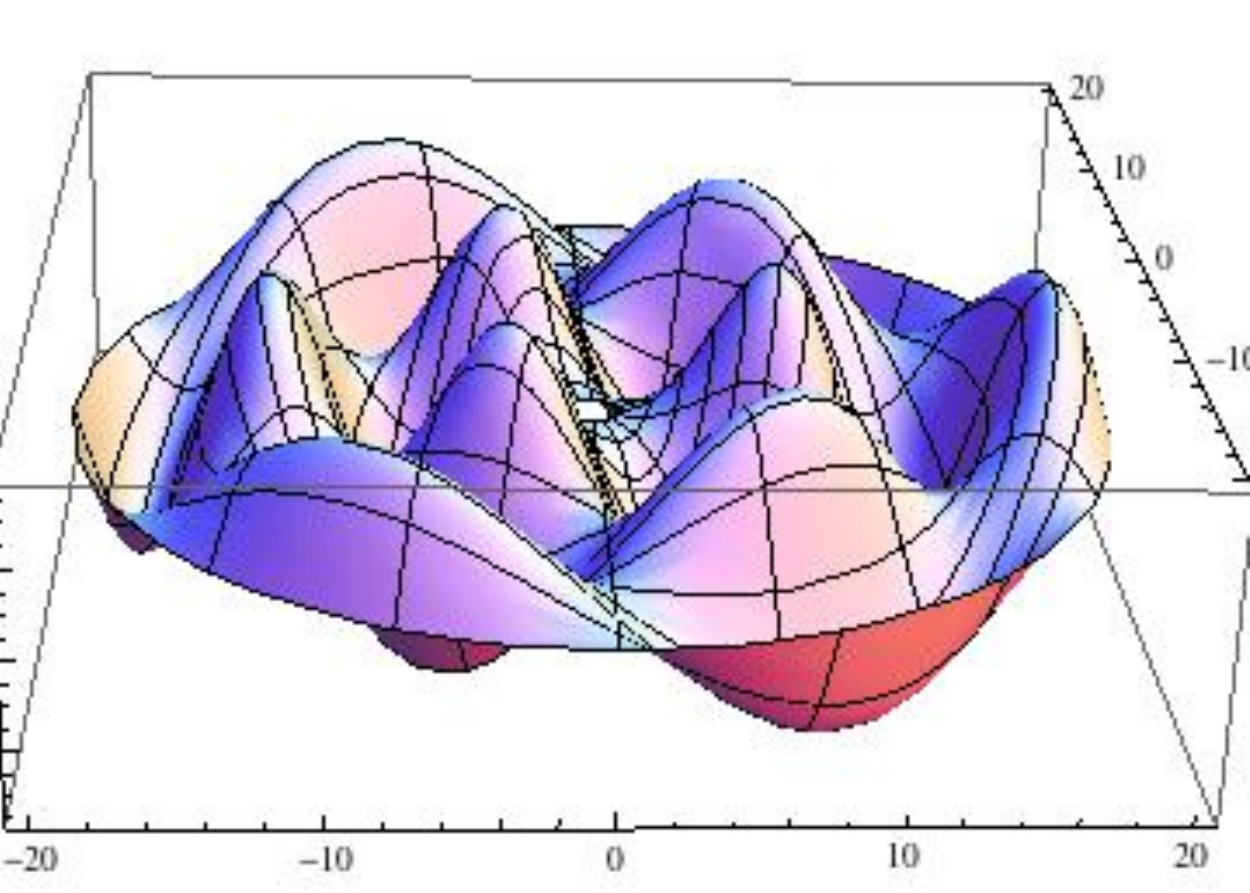} \\ 
\caption{Same as in Fig.~\ref{acoustic_profile1} but for nodeful acoustic clouds with $n=1,m=1$  (top panels)  and  $n=2,m=3$ (bottom panels), on the $r,\phi$ plane. Again, we have fixed $r_0/A=20$ and the cloud supporting background has $B\simeq 0.35$ for the top panels and $B\simeq 0.22$ for the bottom ones.}
\label{acoustic_profile2}
\end{figure}

\begin{figure}[h!]
\centering
\includegraphics[height=2.6in]{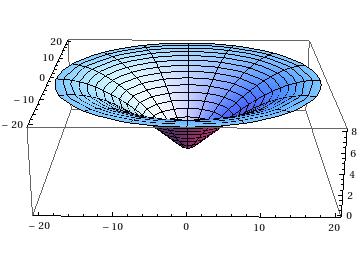}
\includegraphics[height=2.45in]{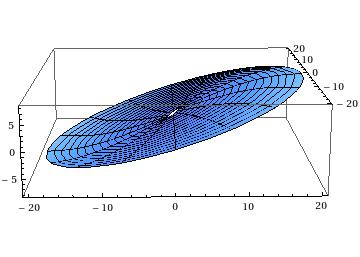} \\ 
\includegraphics[height=2.6in]{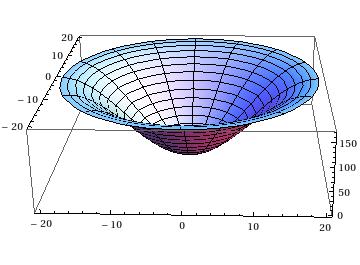}
\includegraphics[height=2.45in]{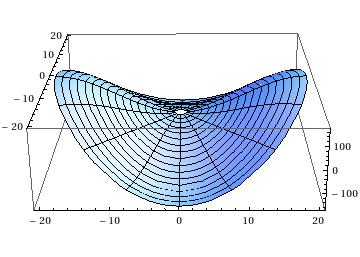} \\ 
\caption{Same as in Fig.~\ref{acoustic_profile1} but for Neumann boundary conditions. Again, we have fixed $r_0/A=20$ and the cloud supporting background has $B\simeq 0.11$ for the top panels and $B\simeq 0.08$ for the bottom ones.}
\label{acoustic_profile3}
\end{figure}

\begin{figure}[h!]
\centering
\includegraphics[height=2.6in]{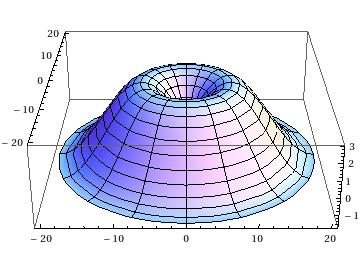}
\includegraphics[height=2.45in]{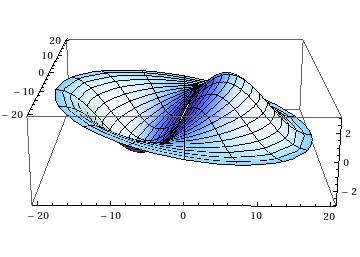} \\ 
\includegraphics[height=2.6in]{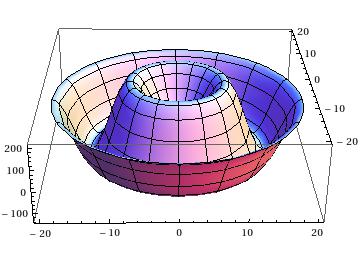}
\includegraphics[height=2.45in]{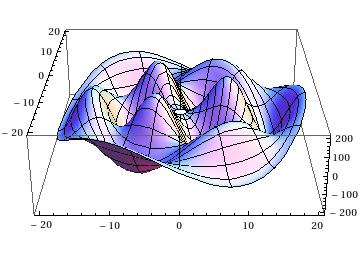} \\ 
\caption{Same as in Fig.~\ref{acoustic_profile2} but for Neumann boundary conditions ($i.e.$ nodeful acoustic clouds with $n=1,m=1$  (top panel)  and  $n=2,m=3$ (bottom panel), on the $r,\phi$ plane). Again, we have fixed $r_0/A=20$ and the cloud supporting background has $B\simeq 0.27$ for the top panels and $B\simeq 0.19$ for the bottom ones.}
\label{acoustic_profile4}
\end{figure}

\end{widetext}

%%%%%%%%%%%%%%
\section{Conclusion}
\label{sec_conclusions}
%%%%%%%%%%%%%%%
The connection between the perturbations of a Newtonian fluid flow and the relativistic physics described by a curved metric -- the acoustic spacetime -- is a surprising and remarkable one. This connection between what seems to be \textit{a priori} two very different worlds can be useful both ways: to use the technology of Lorentzian geometry to learn about hydrodynamics and to use (experimentally controlable) fluids to learn about General Relativity and, in particular, BH physics. In this paper we have been concerned about the former viewpoint and have added yet another property seen in BH physics to the list that can be mimicked by a BH acoustic analogue: the existence of standing waves of a scalar field around a rotating BH; these take the form of \textit{acoustic clouds} around a BH analogue. 

One of the insights that may be obtained from studying analogue models is to separate phenomena that are intrinsic to General Relativity from those that occur whenever Lorentzian geometry may be involved. The example we have provided shows that the existence of stationary waves of a scalar field in the presence of an event horizon is not unique to the BHs of General Relativity and suggests they occur for a large class of BH analogues described by an appropriate Lorentzian geometry, regardless of the dynamical equations obeyed by the background. On the other hand, the equations that rule the background are crucial to promote the scalar clouds seen around Kerr BHs to scalar hairy BHs~\cite{Herdeiro:2014goa}. This is the point at which the parallelism between the BHs from General Relativity and its acoustic analogues is likely to end: whereas for the former Kerr BHs with scalar hair exist, continuously connecting to the Kerr BHs that support scalar clouds, it is unclear if any notion of a deformed acoustic hole by backreacting acoustic clouds can be defined.  

Finally, it is an exciting possibility that the acoustic clouds described herein could be seen in a laboratory experiment. Experimental research on analogue systems is ongoing, but still in its infancy (see~\cite{Richartz:2014lda} for a list of references). We hope the result described here can provide a further research direction for these experiments.\footnote{After the completion of this work, S. Hod posted a paper in the arXiv where the Kerr system with a mirror is studied~\cite{Hod:2014pza}. The main focus therein is the minimum mirror radius which allows the existence of clouds. The general trend observed in~\cite{Hod:2014pza} agrees with our results, whereas the (semi-analytic) technique used therein applies only to the special case of fast spinning Kerr BHs.}

%%%%%%%%%%%%%%%%%%%%%%%%%%%%%%%%%%%%%%%%%%%%%%%
\section*{Acknowledgements}
%%%%%%%%%%%%%%%%%%%%%%%%%%%%%%%%%%%%%%%%%%%%%%%
We would like to thank Juan Carlos Degollado for comments on a draft of this paper. The work in this paper is supported by the FCT Investigator program, grants PTDC/FIS/116625/2010,  NRHEP--295189-FP7-PEOPLE-2011-IRSES and the CIDMA strategic funding UID/MAT/04106/2013. The authors would like also to thank Conselho Nacional de Desenvolvimento Cient\'ifico e Tecnol\'ogico (CNPq), Coordena\c{c}\~ao de Aperfei\c{c}oamento de Pessoal de N\'ivel Superior (CAPES), and Funda\c{c}\~ao Amaz\^onia Paraense de Amparo \`a Pesquisa (FAPESPA), from Brazil, for partial financial support.

\newpage

\bibliographystyle{h-physrev4}
\bibliography{clouds_box}

\end{document}